\documentclass[12pt]{article}
\usepackage{graphicx,amsmath,mathtools}
\usepackage{hyperref}
\usepackage{bm,makecell}
\usepackage{siunitx}
\usepackage{color}
\usepackage{bold-extra}
\usepackage{multirow}
\usepackage[T1]{fontenc}

\newcommand{\codeversion}{3.00}
\newcommand{\hfbtho}{{\sc hfbtho}}
\newcommand{\hfodd}{{\sc hfodd}}
\newcommand{\hfbrad}{{\sc hfbrad}}

\definecolor{db}{rgb}{0.180,0.543,0.340}

\newcounter{mysubsubsection}
\newcounter{leteq}

\newcommand{\hodd}{\bf\color{red}}
\newcommand{\htho}{\bf\color{blue}}
\newcommand{\spc}{{\ }}
\newcommand{\pr}[1]{{\sc{\lowercase{#1}}}}

\newcommand{\nn}{\boldsymbol{n}}

\newcommand{\qvec}{\boldsymbol{q}}
\newcommand{\lc}{\ell_{\text{cut}}(\gras{r})}
\newcommand{\kc}{k_{\text{cut}}(\gras{r})}
\newcommand{\kF}{k_{F}(\gras{r})}
\newcommand{\Mstar}{M^{*}(\gras{r})}
\newcommand{\gras}[1]{\boldsymbol{#1}}
\newcommand{\tensor}[1]{\mathsf{#1}}

\newcommand{\bnl}{\begin{eqnalpha}}
\newcommand{\enl}{\end{eqnalpha}}
\newcommand{\bnll}[1]{\begin{eqnalphalabel}{#1}}
\newcommand{\enll}{\end{eqnalphalabel}}

\newcommand{\keyw}{{\bf Keyword:}}
\newcommand{\key}[1]{\vspace{1ex}\noindent\keyw{\spc}{\tk{#1}}
                         \newline\phantom{\keyw{\spc}{\tk{#1}}}{\spc}}

\newcommand{\be}{\begin{equation}}
\newcommand{\ee}{\end{equation}}
\newcommand{\ba}{\begin{array}}
\newcommand{\ea}{\end{array}}
\newcommand{\bn}{\begin{eqnarray}}
\newcommand{\en}{\end{eqnarray}}
\newcommand{\bc}{\begin{center}}
\newcommand{\ec}{\end{center}}
\newcommand{\bi}{\begin{itemize}}
\newcommand{\ei}{\end{itemize}}

\newcommand{\tv}[1]{{\tt{#1}}\index{#1}}
\newcommand{\tk}[1]{{\tt{#1}}\index{#1}}

\newcommand{\tf}[1]{{\tt{#1}}\index{#1}}

\renewcommand{\tv}[1]{\textcolor{red}    {{\tt{#1}}}}
\renewcommand{\tk}[1]{\textcolor{blue}   {{\tt{#1}}}}

\renewcommand{\tf}[1]{\textcolor{magenta}{{\tt{#1}}}}
\renewcommand{\tv}[1]{{\tt{#1}}}
\renewcommand{\tk}[1]{{\tt{#1}}}

\renewcommand{\tf}[1]{{\tt{#1}}}

\newenvironment{eqnalpha}{\setcounter{leteq}{1}

\begin{eqnarray}}{\end{eqnarray}%
}

\newenvironment{eqnalphalabel}[1]{\setcounter{leteq}{1}
\raisebox{0cm}[0cm][0cm]{\begin{minipage}{1cm}%
\begin{eqnarray}\label{#1}&&\nonumber\end{eqnarray}\end{minipage}}

\begin{eqnarray}}{\end{eqnarray}%
}

\begin{document}

\vspace{0.5cm}
\begin{center}
        {\bf\Large
                     Axially deformed solution of the Skyrme-Hartree-Fock-Bogolyubov equations using the transformed harmonic oscillator basis \\[1ex]
                     (III) \textsc{hfbtho} (v{\codeversion}): a new version of the program.
        }

\vspace{5mm}
        {\large
                       R. Navarro Perez,$^{a}$
                       N. Schunck,$^{a}$\footnote
                       {E-mail: schunck1@llnl.gov}
                       R.-D. Lasseri,$^{b}$
                       C. Zhang,$^{c}$
                       J. Sarich,$^{d}$
        }

\vspace{3mm}
        {\it
          $^a$Nuclear and Chemical Science Division, Lawrence Livermore National Laboratory, Livermore, CA 94551, USA \\
          $^b$Institut de Physique Nucl\'eaire, IN2P3-CNRS, Universit\'e  Paris-Sud, Universit\'e Paris-Saclay, F-91406 Orsay Cedex, France \\
          $^c$NSCL/FRIB Laboratory, Michigan State University, East Lansing, MI 48824, USA \\
          $^d$Mathematics and Computer Science Division, Argonne National Laboratory, Argonne, IL 60439, USA \\
        }
\end{center}

\vspace{5mm}
\hrule

\vspace{2mm}
\noindent{\bf Abstract}

We describe the new version {\codeversion} of the code \pr{hfbtho} that solves
the nuclear Hartree-Fock (HF) or Hartree-Fock-Bogolyubov (HFB)
problem by using the cylindrical transformed deformed harmonic oscillator basis.
In the new version, we have implemented the following features:
(i)   the full Gogny force in both particle-hole and particle-particle channels,
(ii)  the calculation of the nuclear collective inertia at the perturbative
      cranking approximation,
(iii) the calculation of fission fragment charge, mass and deformations based
      on the determination of the neck
(iv)  the regularization of zero-range pairing forces
(v)   the calculation of localization functions
(vi)  a MPI interface for large-scale mass table calculations.
\vspace{2mm}
\hrule

\vspace{2mm}

\vspace{5mm}
{\bf\large NEW VERSION PROGRAM SUMMARY}

\bigskip\noindent{\it Title of the program:} \pr{hfbtho}
                  v{\codeversion}

\bigskip\noindent{\it Catalogue number:}
                   ....

\bigskip\noindent{\it Program obtainable from:}
                      CPC Program Library, \
                      Queen's University of Belfast, N. Ireland
                      (see application form in this issue)

\bigskip\noindent{\it Reference in CPC for earlier version of program:}
                      M.V. Stoitsov, N. Schunck, M. Kortelainen, N. Michel, H. Nam,
                      E. Olsen, J. Sarich, and S. Wild,
                      Comput.\ Phys.\ Commun.\ {\bf 184} (2013).

\bigskip\noindent{\it Catalogue number of previous version:}
                      ADUI\_v2\_0

\bigskip\noindent{\it Licensing provisions:} GPL v3

\bigskip\noindent{\it Does the new version supersede the previous one:} Yes

\bigskip\noindent{\it Computers on which the program has been tested:}
                      Intel Xeon, AMD-Athlon, AMD-Opteron, Cray XT5, Cray XE6

\bigskip\noindent{\it Operating systems:} UNIX, LINUX

\bigskip\noindent{\it Programming language used:} FORTRAN-95

\bigskip\noindent{\it Memory required to execute with typical data:} 200 Mwords

\bigskip\noindent{\it No. of bits in a word:} 8

\bigskip\noindent{\it Has the code been vectorised?:} Yes

\bigskip\noindent{\it Has the code been parallelized?:} Yes

\bigskip\noindent{\it No.{\spc}of lines in distributed program:}
                      20 458

\bigskip\noindent{\it Keywords:}
                      Nuclear many-body problem; Density functional theory; Energy density 
                      functional theory; Self-consistent mean field; Hartree-Fock-Bogolyubov 
                      theory; Finite-temperature HArtree-Fock-Bogolyubov theory; Skyrme 
                      interaction; Gogny force; Pairing correlations; Particle number 
                      projection; Pairing regularization; Collective inertia; Nuclear matter; 
                      Constrained calculations; Potential energy surface; Harmonic oscillator;
                      Transformed harmonic oscillator; Shared memory parallelism; Distributed 
                      memory parallelism; Tensor contractions; Loop optimization.

\bigskip\noindent{\it Nature of physical problem}

\noindent
{\hfbtho} is a physics computer code that is used to model the structure of the
nucleus. It is an implementation of the energy density functional (EDF) approach
to atomic nuclei, where the energy of the nucleus is obtained by integration over 
space of some phenomenological energy density, which is itself a functional of 
the neutron and proton intrinsic densities. In the present version of {\hfbtho}, 
the energy density derives either from the zero-range Skyrme or the finite-range 
Gogny effective two-body interaction between nucleons. Nuclear superfluidity is 
treated at the Hartree-Fock-Bogoliubov (HFB) approximation. Constraints on the 
nuclear shape allows probing the potential energy surface of the nucleus as needed 
e.g., for the description of shape isomers or fission. The implementation of a 
local scale transformation of the single-particle basis in which the HFB solutions 
are expanded provide a tool to properly compute the structure of weakly-bound 
nuclei.

\bigskip\noindent{\it Method of solution}

\noindent
The program uses the axial Transformed Harmonic Oscillator (THO) single-particle
basis to expand quasiparticle wave functions. It iteratively diagonalizes the
Hartree-Fock-Bogolyubov Hamiltonian based on generalized Skyrme-like energy
densities and zero-range pairing interactions or the finite-range Gogny force
until a self-consistent solution is found. A previous version of the program was
presented in M.V. Stoitsov, N. Schunck, M. Kortelainen, N. Michel, H. Nam,
E. Olsen, J. Sarich, and S. Wild, Comput.\ Phys.\ Commun.\ {\bf 184} (2013)
1592--1604 with much of the formalism presented in the original paper
M.V. Stoitsov, J. Dobaczewski, W. Nazarewicz, P. Ring, Comput.\ Phys.\ Commun.\ 
{\bf 167} (2005) 43--63.

\bigskip\noindent{\it Summary of revisions}

\noindent
\begin{enumerate}
\setlength{\itemsep}{-1ex}
\item the Gogny force in both particle-hole and particle-particle channels was
      implemented;
\item the nuclear collective inertia at the perturbative cranking approximation
      was implemented;
\item fission fragment charge, mass and deformations were implemented based
      on the determination of the position of the neck between nascent fragments;
\item the regularization method of zero-range pairing forces was implemented;
\item the localization functions of the HFB solution were implemented;
\item a MPI interface for large-scale mass table calculations was implemented.
\end{enumerate}

\bigskip\noindent{\it Restrictions on the complexity of the problem}

\noindent
Axial- and time-reversal symmetries are assumed.

\bigskip\noindent{\it Typical running time}

\noindent
Highly variable, as it depends on the nucleus, size of the basis, requested
accuracy, requested configuration, compiler and libraries, and hardware
architecture. An order of magnitude would be a few seconds for ground-state
configurations in small bases $N_{\text{max}}\approx 8-12$, to a few minutes
in very deformed configuration of a heavy nucleus with a large basis
$N_{\text{max}}> 20$.

\bigskip\noindent{\it Unusual features of the program}

\noindent
The user must have access to (i) the LAPACK subroutines \pr{DSYEEVR}, \pr{DSYEVD}, \pr{DSYTRF}
and \pr{DSYTRI}, and their dependencies, which compute eigenvalues and
eigenfunctions of real symmetric matrices, (ii) the LAPACK subroutines \pr{DGETRI}
and \pr{DGETRF}, which invert arbitrary real matrices, and (iii) the BLAS routines
\pr{DCOPY}, \pr{DSCAL}, \pr{DGEMM} and \pr{DGEMV} for double-precision linear
algebra (or provide another set of subroutines that can perform such tasks). The
BLAS and LAPACK subroutines can be obtained from the Netlib Repository at the
University of Tennessee, Knoxville: \verb+http://netlib2.cs.utk.edu/+.

\bigskip

{\bf\large LONG WRITE-UP}

\bigskip


\section{Introduction}
\label{sec:intro}

The program {\hfbtho} solves the Hartree-Fock-Bogolyubov (HFB) equation for 
atomic nuclei by expanding the solution on the harmonic oscillator basis. 
The original version 1.66 of the program published in \cite{stoitsov2005} 
had the following characteristics: the HFB equation was only solved for 
even-even nuclei with Skyrme two-body effective potentials in the 
particle-hole channel and contact pairing interactions; the code 
implemented the transformed harmonic oscillator basis and particle number 
projection; the only available constraint was on the axial quadrupole deformation 
and parity-conservation of the HFB solution was explicitly enforced. The 
release 2.00d published in \cite{stoitsov2013} removed some of these 
restrictions and added several new capabilities: parity-breaking of the 
HFB solution was implemented, constraints on axial multipole moments from 
$\lambda = 2$ to $\lambda =9$ became available and the method to handle these 
constraints was based on the readjustment of the Lagrange parameters based 
on the cranking approximation of the QRPA matrix; odd nuclei could be computed 
based on the equal filling approximation of the blocking prescription, and 
the code also implemented the finite-temperature extension of the HFB 
equation.

This release introduces several major features in the code. Most notably, 
version {\codeversion} of {\hfbtho} can now solve the HFB equation for the 
finite-range Gogny force, in both the particle-hole and particle-particle 
channel. The new version also allows the computation of the collective 
inertia tensor and zero-point energy correction within the generator 
coordinate method (GCM) and adiabatic time-dependent Hartree-Fock-Bogolyubov 
(ATDHFB) approximations. It also contains a small fission toolkit that 
gives the charge, mass and deformations of the fission fragments based on 
position of the neck between pre-fragments -- defined as the expectation value of the Gaussian 
neck operator. Finally, the code also implements the regularization of the 
pairing interaction and the calculation of localization functions.

In section \ref{sec:modifs}, we review the modifications introduced in this 
version. In section \ref{sec:benchmarks}, we give a few numerical 
benchmarks for the new capabilities. Finally, in section \ref{sec:input}, 
we discuss the new options available in the input file and explain how to 
run the code, in particular in parallel mode.


\section{Modifications introduced in version {\codeversion}}
\label{sec:modifs}

\setcounter{mysubsubsection}{0}

We present in this section the major new features added to the code between
version 2.00d and \codeversion. 


\subsection{Gogny force}
\label{subsec:gogny}

Version {\codeversion} of {\hfbtho} implements energy functionals 
derived from the Gogny two-body effective interaction~\cite{decharge1980},
\begin{equation}
  \begin{aligned}
    \hat{V}(\bm{r}_1,\bm{r}_2) &= \sum_{i=1,2}
    e^{-(\bm{r}_1-\bm{r}_2)^2/\mu_i^2} \left(W_i + B_i \hat{P}_\sigma
    -H_i\hat{P}_\tau - M_i \hat{P}_\sigma \hat{P}_\tau \right)  \\
    & +  t_0 \left(1+x_0\hat{P}_\sigma \right)\rho^\alpha
    \left(\frac{\bm{r}_1+\bm{r}_2}{2} \right) \delta( \bm{r}_1-\bm{r}_2) \\
    & +  i W_{LS} (\bm{\sigma}_1 + \bm{\sigma}_2) \cdot
    (\overleftarrow{\nabla}_1 - \overleftarrow{\nabla}_2) \times
    \delta(\bm{r}_1-\bm{r}_2) (\overrightarrow{\nabla}_1 -
    \overrightarrow{\nabla}_2), 
  \end{aligned}
  \label{eq:GognyPotential}
\end{equation}
where $\hat{P}_\sigma$ and $\hat{P}_\tau$ are the spin and isospin
exchange operators. The D1, D1S, D1' and D1N parametrizations of the
Gogny force are implemented in this version; see ~\cite{decharge1980,
  berger1991, chappert2008} for information about each of these
parametrizations. Calculations can be performed using any of these
parametrizations by setting the functional name in the {\tt
  hfbtho\_NAMELIST.dat} file to {\tt D1}, {\tt D1S}, {\tt D1p} or
{\tt D1N}.

In previous versions of {\hfbtho} the HFB fields in coordinate space 
were directly integrated using Gaussian quadrature at each iteration to calculate the HFB
matrix. Given the contact nature of
the Skyrme-like functionals that had been implemented so far, the
precise integration of the HFB fields requires only a few dozens of
points. The same level of precision for the finite range part in the
Gogny functional would require several hundreds of points, which
would significantly increase the computation time at each
iteration. To avoid this bottleneck the finite-range contributions to
the HFB fields are computed in configuration space, that is, by
performing the following contractions of the matrix elements of the
anti-symmetrized potential with either the density matrix $\rho$ or 
the pairing tensor $\kappa$
\begin{equation}
\Gamma_{\nn_1 \nn_3} 
= \sum_{\nn_2\nn_4} \langle \nn_1 \nn_2 | \hat{V} \hat{\mathcal{A}} | \nn_3 \nn_4 \rangle 
\rho_{\nn_4\nn_2}, 
\quad 
\Delta_{\nn_1\nn_2} 
= \frac{1}{2} \sum_{\nn_3\nn_4} \langle \nn_1 \nn_2 | \hat{V} \hat{\mathcal{A}} | \nn_3 \nn_4\rangle 
\kappa_{\nn_3 \nn_4}.
  \label{eq:GammaDelta}
\end{equation}
Note that in these expressions, summations extend over the complete 
basis of the single-particle Hilbert space, that is, the basis made of 
the states $(|\nn\rangle,|\bar{\nn}\rangle)$, where $|\bar{\nn}\rangle 
= \hat{T}|\nn\rangle$. In this basis, the matrix of the pairing tensor 
obeys $\kappa_{\nn_3 \nn_4} = \kappa_{\bar{\nn}_3 \bar{\nn}_{4}} = 0$, 
as a consequence of the particular structure of the Bogoliubov matrices; 
see section \ref{subsec:inertia}. In practice, {\hfbtho} uses what is 
known as the russian convention in the pairing channel, that is, it 
works with the pairing density $\tilde{\rho}$ instead of the pairing 
tensor $\kappa$ \cite{dobaczewski1984}. In configuration space, the two 
are related through 
\begin{equation}
\tilde{\rho}_{\nn_3\nn_4} = -2\sigma_{\nn_4}\kappa^{*}_{\nn_3 \bar{\nn}_4}.
\end{equation}
As a consequence, it is more convenient to also use the Hermitian pairing 
field $\tilde{h}$ instead of the antisymmetric pairing field $\Delta$,
\begin{equation}
\tilde{h}_{\nn_1\nn_2} = -2\sigma_{\nn_2}\Delta^{*}_{\nn_1 \bar{\nn}_2}.
\end{equation}
An additional simplification specific to built-in time-reversal symmetry 
in {\hfbtho} is that the pairing tensor, density and pairing fields are 
real and symmetric \cite{chappert2015}. Therefore, we will use the 
following formula in the pairing channel,
\begin{equation}
\tilde{h}_{\nn_1\nn_2} 
= \frac{1}{2} \sum_{\nn_3\nn_4} \langle \nn_1 \bar{\nn}_2 | \hat{V} \hat{\mathcal{A}} | \nn_3 \bar{\nn}_4\rangle 
\sigma_{\nn_2} \sigma_{\nn_4} \tilde{\rho}_{\nn_4 \nn_3}.
\end{equation}
In these expressions, the basis states $|\nn\rangle$ are defined by
the quantum numbers of the HO, $\nn \equiv (n_r, \Lambda, n_z, \sigma_{\gras{n}},\tau_{\gras{n}})$  
with the spin projection $\Sigma_{\nn} = \sigma_{\nn}/2 = \pm 1/2$
and isospin $\tau_{\nn}$ quantum numbers; $\hat{\mathcal{A}} = 1 -
\hat{P}_r \hat{P}_\sigma \hat{P}_\tau $ is the anti-symmetrization
operator; under time-reversal operation, $\Lambda$ and $\sigma_{\nn}$ 
change sign, hence, $\nn \equiv (n_r, \Lambda, n_z, \sigma_{\gras{n}},\tau_{\gras{n}}) 
\Rightarrow \bar{\nn} \equiv (n_r, -\Lambda, n_z, -\sigma_{\gras{n}},\tau_{\gras{n}})$.
Finally, note that the matrix elements of the two-body potential are 
independent of the HFB iterations. This allows computing them only 
once at the beginning of the iterative loop.

\subsubsection{Matrix element of a two-body Gaussian potential}

The calculation of the finite-range part of the full two-body matrix
element of potential (\ref{eq:GognyPotential})
requires evaluating spatial integrals of the type
\begin{align}
V_{\nn_1\nn_2\nn_3\nn_4}
& = \langle \nn_1\nn_2 | e^{-(\bm{r}_1-\bm{r}_2)^2/\mu^2} | \nn_3\nn_4 \rangle  \\ 
& =  \int d^3\gras{r}_1 \int d^3\gras{r}_2\;  
\phi^*_{\nn_1}(\bm{r}_1) \phi^*_{\nn_2}(\bm{r}_2) 
e^{-(\bm{r}_1-\bm{r}_2)^2/\mu^2} 
\phi_{\nn_3}(\bm{r}_1) \phi_{\nn_4}(\bm{r}_2) ,
\label{eq:GaussianMatrixElement}
\end{align}
where $\phi_{\nn}(\bm{r})$ is the spatial part of the stretched 
harmonic oscillator basis function in the cylindrical coordinates 
$\gras{r} \equiv (\rho,z,\varphi)$. A convenient property of the 
two-body Gaussian potential and the HO wave function is that each 
of them can be separated into a product of axial and polar 
functions, namely
\begin{equation}
e^{-(\bm{r}_1-\bm{r}_2)/\mu^2} 
=
e^{-(\bm{\rho}_1-\bm{\rho}_2)/\mu^2} e^{-(z_1-z_2)/\mu^2}, 
\quad
\phi_{\nn}(\bm{r}) = \phi_{n_r}^{\Lambda}(\rho,\varphi) \phi_{n_z}(z).
\end{equation}
This allows separating the matrix element 
(\ref{eq:GaussianMatrixElement}) into
\begin{equation}
V_{\nn_1\nn_2\nn_3\nn_4} = 
V_{n_r^{(1)} \Lambda^{(1)}, n_r^{(2)} \Lambda^{(2)}, 
   n_r^{(3)} \Lambda^{(3)}, n_r^{(4)} \Lambda^{(4)}}
V_{n_z^{(1)}n_z^{(2)}n_z^{(3)}n_z^{(4)}}.
\end{equation}

Our implementation to calculate these matrix elements closely follows 
the method outlined in~\cite{younes2009-b}. For the axial matrix
element, we use the expansion
\begin{equation}
  V_{n_z^{(1)}n_z^{(2)}n_z^{(3)}n_z^{(4)}} =
  \frac{\mu}{\sqrt{2\pi^3}b_z}
  \sum_{n=|n_z^{(2)}-n_z^{(4)}|,2}^{n_z^{(2)}+n_z^{(4)}} T_{n_z^{(2)}
    n_z^{(4)}}^{n}\bar{F}_{n_z^{(1)} n_z^{(3)}}^{n},
\label{eq:hyper_axial}
\end{equation}
which preserves numerical precision at large number of harmonic
oscillator shells. The coefficient $T_{n_z^{(2)}n_z^{(4)}}^{n}$ is
defined as
\begin{equation}
  T_{n_z^{(2)}n_z^{(4)}}^{n} \equiv \frac{\sqrt{n_z^{(2)}! n_z^{(4)}!
      n!}}{\left(\frac{n_z^{(2)} -n_z^{(4)} +n}{2}\right)!
    \left(\frac{n_z^{(4)} -n_z^{(2)} +n}{2}\right)!
    \left(\frac{n_z^{(2)} +n_z^{(4)} -n}{2}\right)!},
\end{equation}
and the function $\bar{F}_{n_z^{(1)} n_z^{(3)}}^{n}$ as
\begin{equation}
\bar{F}_{n_z^{(1)} n_z^{(3)}}^{n} 
\equiv
\frac{\Gamma(\xi-n_z^{(1)}) \Gamma(\xi-n_z^{(3)})
\Gamma(\xi-n)}{\left(1 + \frac{\mu^2}{2b_z^2} \right)^\xi 
\sqrt{n!n_z^{(1)}! n_z^{(3)}!}} 
{}_2F_1 \Big(-n_z^{(1)}, -n_z^{(3)}; -\xi +
  n + 1; -\frac{\mu^2}{2b_z^2}\Big),
\end{equation}
where $\xi = (n_z^{(1)} + n_z^{(3)} + n + 1)/2$, $_2F_1$ is the
hyper-geometric function and $\Gamma$ is the usual gamma function.

Since there is no analogous expansion to (\ref{eq:hyper_axial}) for the 
polar matrix element, we expand the two-dimensional HO wave function into 
a sum of products of two one-dimensional HO functions (equivalent to a 
change from polar to Cartesian coordinates),
\begin{equation}
\phi_{n_r}^{\Lambda}(\rho,\varphi) 
= \sum_{n_y=0}^{2n_r + |\Lambda|} C_{n_x n_y}^{n_r \Lambda} 
\phi_{n_x}(x) \phi_{n_y}(y),
\end{equation}
where $n_x = 2n_r + |\Lambda| - n_y$ and
\begin{equation}
C_{n_x n_y}^{n_r \Lambda} 
\equiv \frac{2^{-n_r- |\Lambda|/2}
(-1)^{n_r} \sqrt{n_r!(n_r+|\Lambda|)!}}{\sqrt{n_x! n_y!} i^{n_y}}
\sum_{q=0}^{\min\left(n_y,n_r+\frac{|\Lambda|-\Lambda}{2}\right)} 
{{n_x}\choose{n-q + \frac{|\Lambda|-\Lambda}{2}}} {{n_y}\choose{q}} (-1)^{n_y-q}.
\end{equation}
This expansion transforms the polar matrix elements into a sum of
products of axial matrix elements, namely
\begin{equation}
\begin{aligned}
V_{n_r^{(1)} \Lambda^{(1)}, n_r^{(2)} \Lambda^{(2)}, 
   n_r^{(3)} \Lambda^{(3)} ,n_r^{(4)} \Lambda^{(4)}}
= & 
\sum_{n_y^{(1)}} \sum_{n_y^{(2)}} \sum_{n_y^{(3)}} \sum_{n_y^{(4)}} 
C_{n_x^{(1)}n_y^{(1)}}^{n_r^{(1)}\Lambda^{(1)} *} 
C_{n_x^{(2)}n_y^{(2)}}^{n_r^{(2)}\Lambda^{(2)} *} 
C_{n_x^{(3)}n_y^{(3)}}^{n_r^{(3)}\Lambda^{(3)} } 
C_{n_x^{(4)}n_y^{(4)}}^{n_r^{(4)}\Lambda^{(4)} }  \\
& \times 
V_{n_x^{(1)}n_x^{(2)}n_x^{(3)}n_x^{(4)}}
V_{n_y^{(1)}n_y^{(2)}n_y^{(3)}n_y^{(4)}}.
\label{eq:radialME}
\end{aligned}
\end{equation}
Each axial matrix elements in (\ref{eq:radialME}) can be calculated 
accurately with (\ref{eq:hyper_axial}) by making the substitution 
$b_z \rightarrow b_\bot$. Reference~\cite{younes2009-b} includes other 
simpler and more direct expansions to calculate 
$V_{n_r^{(1)} \Lambda^{(1)}, n_r^{(2)} \Lambda^{(2)}, n_r^{(3)} 
\Lambda^{(3)},n_r^{(4)} \Lambda^{(4)}}$ and 
$V_{n_z^{(1)},n_z^{(2)},n_z^{(3)},n_z^{(4)}}$. These expansions were 
implemented for testing and debugging purposes only. The corresponding 
subroutines and functions are included in version \codeversion but are 
not used, since these expansions loose accuracy as the basis size 
increases. For a complete list of subroutines and functions see the 
documentation for the {\tt hfbtho\_gogny} module.

\subsubsection{Contraction with the two-body matrix elements}

Once the axial and polar matrix elements have been calculated with
(\ref{eq:hyper_axial}) and (\ref{eq:radialME}) before the first
HFB iteration, a contraction with the one body density and pairing
matrices has to be made at each iteration.
  
{\bf Mean field - } For the calculation of the mean field $\Gamma_{\gras{n}_1\gras{n}_3}$ it
is more convenient to apply the anti-symmetrization operator to the
left and embed it into the potential, rather than to the right through
its action on the two-body ket. This results in
\be
\hat{V}_i \hat{\mathcal{A}} 
= V_i(\bm{r})\left(W_i + B_i\hat{P}_\sigma -
H_i\hat{P}_\tau - M_i\hat{P}_\sigma\hat{P}_\tau \right) 
\left(1 - \hat{P}_\sigma\hat{P}_\tau\hat{P}_r\right) 
\equiv \hat{V}_i^{\rm ID} - \hat{V}_i^{\rm IE} \hat{P}_\tau,
\label{eq:AntisymmPotential}
\ee
where the ID and IE labels stand for isospin direct and isospin
exchange respectively.

In this subsection we slightly abuse the notation so that the $\nn_i$
label includes all the quantum numbers that are not written explicitly
or have not been contracted. Inserting (\ref{eq:AntisymmPotential}) 
into the expression for $\Gamma$ in (\ref{eq:GammaDelta}), applying the 
isospin exchange operator to the right and contracting the $\tau$ indices 
(i.e. summing over all possible values of $\tau_{\nn_2}$ and $\tau_{\nn_4}$) 
results in
\begin{equation}
  \begin{alignedat}{1}
    \Gamma_{i,\nn_1\nn_3} = \sum_{\mathclap{\nn_2 \nn_4}} & \Big[ \langle \nn_1 \nn_2 |
      \hat{V}_i^{\rm ID} | \nn_3 \nn_4 \rangle \langle \tau_{\nn_1} | \tau_{\nn_3}
      \rangle ( \rho_{\nn_4 \nn_2}^{\tau\tau} + \rho_{\nn_4 \nn_2}^{\tau'\tau'})  \\
      -&\langle \nn_1 \nn_2 | \hat{V}_i^{\rm IE} | \nn_3 \nn_4 \rangle (\langle
      \tau_{\nn_1} |\tau \rangle \langle \tau| \tau_{\nn_3} \rangle \rho_{\nn_4
        \nn_2}^{\tau\tau} + \langle \tau_{\nn_1} |\tau' \rangle \langle \tau'| \tau_{\nn_3}
      \rangle \rho_{\nn_4 \nn_2}^{\tau'\tau'} ) \Big],
  \end{alignedat}
  \label{eq:GammaIsospin}
\end{equation}
where $\tau \neq \tau'$ and the orthonormality of the isospin states
along with the fact that $\rho$ is block-diagonal in isospin
(i.e. $\rho_{\nn_2 \nn_4}^{\tau\tau'} = 0$) has been used. From
(\ref{eq:GammaIsospin}) one can quickly see that $\Gamma$ is also
block-diagonal in isospin, namely
$\Gamma_{i,\nn_1\nn_3}^{\tau_{\nn_1}\tau_{\nn_3}} = 0$ if $\tau_{\nn_1} \neq
\tau_{\nn_3}$, and removing the second $\tau$ for simplicity we obtain
\begin{align}
\Gamma^{\tau}_{i,\nn_1\nn_3} &= \sum_{\nn_2\nn_4}
\Big[ 
\langle \nn_1 \nn_2 | \hat{V}_i^{\rm ID} - \hat{V}_i^{\rm IE} | \nn_3 \nn_4 \rangle \rho_{\nn_4\nn_2}^{\tau}
+ 
\langle \nn_1 \nn_2 | \hat{V}_i^{\rm ID} | \nn_3 \nn_4 \rangle
\rho_{\nn_4\nn_2}^{\tau'}
\Big], \\
& \equiv \sum_{\nn_2\nn_4} 
\Big[ \langle \nn_1 \nn_2 | \hat{V}_{i,\tau \tau} | \nn_3 \nn_4 \rangle
\rho_{\nn_4\nn_2}^{\tau} 
+ 
\langle \nn_1 \nn_2 | \hat{V}_{i,\tau \tau'} | \nn_3\nn_4 \rangle 
\rho_{\nn_4\nn_2}^{\tau'} 
\Big].
\label{eq:Gammatau}
\end{align}

Separating $\hat{V}_{i,\tau \tau}$ and $\hat{V}_{i,\tau \tau'}$ into spin-direct 
and spin-exchange parts
\begin{equation}
  \begin{alignedat}{2}
    \hat{V}_{i,\tau \tau } &= \hat{V}_{i,\tau \tau }^{\rm SD} + \hat{V}_{i,\tau \tau }^{\rm SE} \hat{P}_\sigma & \quad
    \hat{V}_{i,\tau \tau'} &= \hat{V}_{i,\tau \tau'}^{\rm SD} + \hat{V}_{i,\tau \tau'}^{\rm SE} \hat{P}_\sigma  \\
    \hat{V}_{i,\tau \tau }^{\rm SD} &= V_i(\gras{r}) \left[(W_i-H_i)+(M_i-B_i)\hat{P}_r\right] &
    \hat{V}_{i,\tau \tau'}^{\rm SD} &= V_i(\gras{r}) \left(W_i+M_i\hat{P}_r\right) \\
    \hat{V}_{i,\tau \tau }^{\rm SE} &= V_i(\gras{r}) \left[(B_i-M_i)+(H_i-W_i)\hat{P}_r\right] &
    \hat{V}_{i,\tau \tau'}^{\rm SE} &= V_i(\gras{r}) \left(B_i+H_i\hat{P}_r\right), 
  \end{alignedat}
  \label{eq:Vspin_direct_exchange}
\end{equation}
allows applying the spin exchange operator to the right and contracting
the $\sigma$ indices to obtain
\begin{align}
\Gamma_{i,\nn_1 \nn_3}^\tau = \sum_{\nn_2\nn_4} & 
\Big\{ 
 \langle\sigma_{\nn_1}|\sigma_{\nn_3}\rangle 
 \big[
   \langle \nn_1 \nn_2 |\hat{V}_{i,\tau  \tau}^{\rm SD} | \nn_3 \nn_4 \rangle 
   (\rho_{\nn_4\nn_2}^{\tau  \sigma \sigma} + \rho_{\nn_4\nn_2}^{\tau \sigma' \sigma'}) 
   + 
   \langle \nn_1 \nn_2 |\hat{V}_{i,\tau \tau'}^{\rm SD} | \nn_3 \nn_4 \rangle
   (\rho_{\nn_4\nn_2}^{\tau' \sigma \sigma} + \rho_{\nn_4\nn_2}^{\tau'\sigma' \sigma'} ) 
 \big] \nonumber\\
& 
+ \langle \nn_1 \nn_2| \hat{V}_{i,\tau \tau}^{\rm SE} |\nn_3 \nn_4 \rangle
 \big[
    \langle \sigma_{\nn_1}|\sigma\rangle \langle \sigma |\sigma_{\nn_3}\rangle \rho_{\nn_4\nn_2}^{\tau \sigma \sigma }
   +\langle \sigma_{\nn_1}|\sigma\rangle \langle \sigma'|\sigma_{\nn_3}\rangle \rho_{\nn_4\nn_2}^{\tau \sigma \sigma'} 
   \nonumber\\
   & \phantom{+ \langle \nn_1 \nn_2| \hat{V}_{i,\tau \tau}^{\rm SE} |\nn_3 \nn_4 }
   +\langle \sigma_{\nn_1}|\sigma'\rangle \langle \sigma |\sigma_{\nn_3}\rangle \rho_{\nn_4\nn_2}^{\tau \sigma' \sigma }
   +\langle \sigma_{\nn_1}|\sigma'\rangle \langle \sigma'|\sigma_{\nn_3}\rangle \rho_{\nn_4\nn_2}^{\tau \sigma' \sigma'}
 \big] \\
& + \langle \nn_1 \nn_2| \hat{V}_{i,\tau \tau'}^{\rm SE} |\nn_3 \nn_4 \rangle
 \big[
    \langle \sigma_{\nn_1}|\sigma\rangle \langle \sigma |\sigma_{\nn_3}\rangle \rho_{\nn_4\nn_2}^{\tau' \sigma \sigma }
   +\langle \sigma_{\nn_1}|\sigma\rangle \langle \sigma'|\sigma_{\nn_3}\rangle \rho_{\nn_4\nn_2}^{\tau' \sigma \sigma'} 
   \nonumber \\
   & \phantom{+ \langle \nn_1 \nn_2| \hat{V}_{i,\tau \tau'}^{\rm SE} |\nn_3 \nn_4 }
   +\langle \sigma_{\nn_1}|\sigma'\rangle \langle \sigma |\sigma_{\nn_3}\rangle \rho_{\nn_4\nn_2}^{\tau' \sigma' \sigma }
   +\langle \sigma_{\nn_1}|\sigma'\rangle \langle \sigma'|\sigma_{\nn_3}\rangle \rho_{\nn_4\nn_2}^{\tau' \sigma' \sigma'}
 \big]
\Big\}. \nonumber
\end{align}
This expression gives different results for states with 
$\sigma_{\nn_1}=\sigma_{\nn_3}$, denoted as $\sigma\sigma$, and states
with $\sigma_{\nn_1}\neq\sigma_{\nn_3}$, denoted as $\sigma\sigma'$
\begin{align}
  \Gamma_{i,\nn_1\nn_3}^{\tau \sigma \sigma} 
  & = \sum_{\nn_2\nn_4} 
  \Big[
  \langle \nn_1 \nn_2 | \hat{V}_{i,\tau \tau}^{\rm SD} +  \hat{V}_{i,\tau \tau}^{\rm SE} | \nn_3 \nn_4 \rangle 
  \rho_{\nn_4\nn_2}^{\tau \sigma \sigma} 
  +
  \langle \nn_1 \nn_2 | \hat{V}_{i,\tau \tau}^{\rm SD} | \nn_3 \nn_4 \rangle 
  \rho_{\nn_4\nn_2}^{\tau \sigma' \sigma'} \nonumber \\
  & \quad \quad + \langle \nn_1 \nn_2 | \hat{V}_{i,\tau \tau'}^{\rm SD} + \hat{V}_{i,\tau \tau'}^{\rm SE} | \nn_3 \nn_4 \rangle 
  \rho_{\nn_4\nn_2}^{\tau' \sigma \sigma} +
  \langle \nn_1 \nn_2 | \hat{V}_{i,\tau \tau'}^{\rm SD} | \nn_3 \nn_4 \rangle 
  \rho_{\nn_4\nn_2}^{\tau' \sigma' \sigma'} 
  \Big] \medskip\\
  \Gamma_{i,\nn_1\nn_3}^{\tau \sigma \sigma'} & = \sum_{\nn_2\nn_4} 
  \Big[
    \langle \nn_1\nn_2 | \hat{V}_{i,\tau \tau}^{\rm SE}  | \nn_3 \nn_4 \rangle \rho_{\nn_4\nn_2}^{\tau \sigma \sigma'}
    +
    \langle \nn_1\nn_2 | \hat{V}_{i,\tau \tau'}^{\rm SE} | \nn_3 \nn_4 \rangle \rho_{\nn_4\nn_2}^{\tau' \sigma \sigma'} 
  \Big].
\end{align}

At this point it is important to remember that since axial symmetry is 
explicitly enforced in {\hfbtho}, $\Omega = \Lambda + \sigma/2$ 
is a good quantum number and therefore the $\Gamma$, $\tilde{h}$, $\rho$ and
$\tilde{\rho}$ matrices are block-diagonal. Each block is characterized 
by a different value of $\Omega$. Furthermore, due to time reversal symmetry, 
{\hfbtho} works only with positive values of $\Omega$. However, the summations 
in (\ref{eq:GammaDelta}) include both positive and negative values of 
$\Lambda$ and $\sigma$. Therefore we apply the time reversal operator 
$\hat{T} | \nn \rangle = |\bar{\nn}\rangle $ to blocks with $\Omega<0$ in 
order to express such states in terms of states with $\Omega>0$. This results in
\begin{align}
\Gamma_{i,\nn_1\nn_3}^{\tau \sigma \sigma} 
& = \sum_{\mathclap{\nn_2\nn_4,\Omega>0}} 
\Big[ 
\big(
\langle \nn_1 \nn_2 | \hat{V}_{i,\tau \tau}^{\rm SD} +  \hat{V}_{i,\tau \tau}^{\rm SE} | \nn_3 \nn_4 \rangle 
+
\langle \nn_1 \bar{\nn}_2 | \hat{V}_{i,\tau \tau}^{\rm SD} | \nn_3 \bar{\nn}_4 \rangle
\big) 
\rho_{\nn_4\nn_2}^{\tau \sigma \sigma} \nonumber \\
&\quad \quad + 
\big(
\langle \nn_1 \bar{\nn}_2 | \hat{V}_{i,\tau \tau}^{\rm SD} +  \hat{V}_{i,\tau \tau}^{\rm SE} | \nn_3 \bar{\nn}_4 \rangle 
+
\langle \nn_1 \nn_2 | \hat{V}_{i,\tau \tau}^{\rm SD} | \nn_3 \nn_4 \rangle
\big) \rho_{\nn_4\nn_2}^{\tau \sigma' \sigma'} \nonumber \\
& \quad \quad + 
\big(
\langle \nn_1 \nn_2 | \hat{V}_{i,\tau \tau'}^{\rm SD} +  \hat{V}_{i,\tau \tau'}^{\rm SE} | \nn_3 \nn_4 \rangle 
+
\langle \nn_1 \bar{\nn}_2 | \hat{V}_{i,\tau \tau'}^{\rm SD} | \nn_3 \bar{\nn}_4 \rangle
\big) \rho_{\nn_4\nn_2}^{\tau' \sigma \sigma} 
 \nonumber \\
& \quad \quad + \big(
\langle \nn_1 \bar{\nn}_2 | \hat{V}_{i,\tau \tau'}^{\rm SD} +  \hat{V}_{i,\tau \tau'}^{\rm SE} | \nn_3 \bar{\nn}_4 \rangle 
+
\langle \nn_1 \nn_2 | \hat{V}_{i,\tau \tau'}^{\rm SD} | \nn_3 \nn_4 \rangle) \rho_{\nn_4\nn_2}^{\tau' \sigma' \sigma'} 
\Big], \\
\Gamma_{i,\nn_1\nn_3}^{\tau \sigma \sigma'} 
& = \sum_{\mathclap{\nn_2\nn_4,\Omega>0}} 
\Big[
  \langle \nn_1\nn_2 | \hat{V}_{i,\tau \tau}^{\rm SE} | \nn_3 \nn_4 \rangle \rho_{\nn_4\nn_2}^{\tau \sigma \sigma'} 
  -
  \langle \nn_1 \bar{\nn}_2 | \hat{V}_{i,\tau \tau}^{\rm SE} | \nn_3 \bar{\nn}_4 \rangle \rho_{\nn_4\nn_2}^{\tau \sigma' \sigma} \nonumber \\
  &\quad \quad + \langle \nn_1 \nn_2 | \hat{V}_{i,\tau \tau'}^{\rm SE} | \nn_3 \nn_4 \rangle \rho_{\nn_4\nn_2}^{\tau' \sigma \sigma'} 
  -
  \langle \nn_1 \bar{\nn}_2 | \hat{V}_{i,\tau \tau'}^{\rm SE} | \nn_3 \bar{\nn}_4 \rangle \rho_{\nn_4\nn_2}^{\tau' \sigma' \sigma} 
\Big].
\end{align}

Inserting the definitions (\ref{eq:Vspin_direct_exchange}), it is
possible to separate the direct and exchange contributions to $\Gamma$
due to the presence of $\hat{P}_r$ so that 
$\Gamma_{\nn_1\nn_3}=\Gamma_{\nn_1\nn_3}^{\rm D}+\Gamma_{\nn_1\nn_3}^{\rm E}$.
Strictly speaking this separation is not necessary to
calculate and diagonalize the HFB Hamiltonian, but it provides a convenient 
tool to verify our implementation against other codes. Using the notation 
from (\ref{eq:GaussianMatrixElement}) we find the final result
\begin{align}
\label{eq:GammaDsigma_sigma}  
\Gamma_{i,\nn_1\nn_3}^{{\rm D} \tau \sigma \sigma} 
& = \sum_{\mathclap{\nn_2\nn_4}} 
\Big\{ \big[
   V_{\nn_1\nn_2\nn_3\nn_4}^{(i)} (W_i +B_i -H_i -M_i) + V_{\nn_1\bar{\nn}_2\nn_3\bar{\nn}_4}^{(i)} (W_i-H_i)\big] \rho_{\nn_4\nn_2}^{\tau \sigma \sigma} \nonumber \\
& \quad \quad + \big[
   V_{\nn_1\bar{\nn}_2\nn_3\bar{\nn}_4}^{(i)} (W_i +B_i -H_i -M_i) + V_{\nn_1\nn_2\nn_3\nn_4}^{(i)} (W_i-H_i)\big] \rho_{\nn_4\nn_2}^{\tau \sigma' \sigma'} \nonumber \\
& \quad \quad + \big[
   V_{\nn_1\nn_2\nn_3\nn_4}^{(i)} (W_i +B_i) + V_{\nn_1\bar{\nn}_2\nn_3\bar{\nn}_4}^{(i)} W_i \big] \rho_{\nn_4\nn_2}^{\tau' \sigma \sigma} \nonumber \\
& \quad \quad + \big[
   V_{\nn_1\bar{\nn}_2\nn_3\bar{\nn}_4}^{(i)} (W_i +B_i) + V_{\nn_1\nn_2\nn_3\nn_4}^{(i)} W_i \big] \rho_{\nn_4\nn_2}^{\tau'\sigma' \sigma'}\Big\}, \\
\label{eq:GammaEsigma_sigma}
\Gamma_{i,\nn_1\nn_3}^{{\rm E} \tau \sigma \sigma} 
& = \sum_{\mathclap{\nn_2\nn_4}} 
\Big\{ \big[
   V_{\nn_1\nn_2\nn_4\nn_3}^{(i)} (H_i +M_i -W_i -B_i) + V_{\nn_1\bar{\nn}_2\bar{\nn}_4\nn_3}^{(i)} (M_i-B_i)\big] \rho_{\nn_4\nn_2}^{\tau \sigma \sigma} \nonumber \\
& \quad \quad + \big[
   V_{\nn_1\bar{\nn}_2\bar{\nn}_4\nn_3}^{(i)} (H_i +M_i -W_i -B_i) + V_{\nn_1\nn_2\nn_4\nn_3}^{(i)} (M_i-B_i)\big] \rho_{\nn_4\nn_2}^{\tau \sigma' \sigma'} \nonumber \\
& \quad \quad + \big[
   V_{\nn_1\nn_2\nn_4\nn_3}^{(i)} (H_i +M_i) + V_{\nn_1\bar{\nn}_2\bar{\nn}_4\nn_3}^{(i)} M_i \big] \rho_{\nn_4\nn_2}^{\tau' \sigma \sigma} \nonumber \\
& \quad \quad + \big[
   V_{\nn_1\bar{\nn}_2\bar{\nn}_4\nn_3}^{(i)} (H_i +M_i) + V_{\nn_1\nn_2\nn_4\nn_3}^{(i)} M_i \big] \rho_{\nn_4\nn_2}^{\tau'\sigma' \sigma'}\Big\},  \\
\Gamma_{i,\nn_1\nn_3}^{{\rm D} \tau \sigma \sigma'} 
& = \sum_{\mathclap{\nn_2\nn_4}} 
\Big[(B_i-M_i)(V_{\nn_1\nn_2\nn_3\nn_4}\rho_{\nn_4\nn_2}^{\tau \sigma \sigma'}-V_{\nn_1\bar{\nn}_2\nn_3\bar{\nn}_4}\rho_{\nn_4\nn_2}^{\tau \sigma' \sigma}) \nonumber \\
& \quad \quad + 
  B_i (V_{\nn_1\nn_2\nn_3\nn_4}\rho_{\nn_4\nn_2}^{\tau' \sigma \sigma'}-V_{\nn_1\bar{\nn}_2\nn_3\bar{\nn}_4}\rho_{\nn_4\nn_2}^{\tau' \sigma' \sigma})\Big], \\
\label{eq:GammaEsigma_sigmap}
\Gamma_{i,\nn_1\nn_3}^{{\rm E} \tau \sigma \sigma'} 
& = \sum_{\mathclap{\nn_2\nn_4}} 
\Big[(H_i-W_i)(V_{\nn_1\nn_2\nn_4\nn_3}\rho_{\nn_4\nn_2}^{\tau \sigma \sigma'}-V_{\nn_1\bar{\nn}_2\bar{\nn}_4\nn_3}\rho_{\nn_4\nn_2}^{\tau \sigma' \sigma}) \nonumber \\
& \quad \quad + 
  H_i (V_{\nn_1\nn_2\nn_4\nn_3}\rho_{\nn_4\nn_2}^{\tau' \sigma \sigma'}-V_{\nn_1\bar{\nn}_2\bar{\nn}_4\nn_3}\rho_{\nn_4\nn_2}^{\tau' \sigma' \sigma})\Big],
\end{align}
where the $\Omega>0$ label in the summations is implicit. Finally, we
note that $V_{\nn_1\nn_2\nn_3\nn_4} =
V_{\nn_1\bar{\nn}_2\nn_3\bar{\nn}_4}$ along with
$\rho_{\nn_4\nn_2}^{\tau\sigma\sigma'} =
\rho_{\nn_4\nn_2}^{\tau\sigma'\sigma}$ results in
$\Gamma_{i,\nn_1\nn_3}^{{\rm D} \tau \sigma \sigma'} = 0$.

{\bf Pairing field $\gras{\tilde{h}}$ - } For the contraction of the
two-body matrix elements with the pairing density $\tilde{\rho}$, we
take advantage of the fact that for local two-body potentials, the
exchange and direct contributions to the pairing field are
identical. This cancels out the $1/2$ factor and only the direct part
needs to be calculated, i.e.,
\begin{equation}
\tilde{h}_{\nn_1\nn_2} = \sum_{\nn_3\nn_4}
  \langle \nn_1 \bar{\nn}_2 | \hat{V} | \nn_3 \bar{\nn}_4\rangle 
  \sigma_{\nn_2} \sigma_{\nn_4} \tilde{\rho}_{\nn_4 \nn_3}.
\end{equation}

As with $\Gamma$, after contracting the $\tau$ indices $\tilde{h}$
turns out to be block diagonal in isospin since $\tilde{\rho}$ is also
block diagonal in isospin and the states $|\tau\rangle$ are orthonormal,
\begin{equation}
  \tilde{h}_{i,\nn_1\nn_2}^{\tau} = \sum_{\nn_3\nn_4} \langle \nn_1 \bar{\nn}_2 |
  \hat{V}_i(\gras{r}) \big[ W_i - H_i + (B_i-M_i)\hat{P}_\sigma \big] | \nn_3 \bar{\nn}_4
  \rangle \sigma_{\nn_2} \sigma_{\nn_4} \tilde{\rho}_{\nn_4\nn_3}^\tau.
\end{equation}

Now, the contraction with respect to $\sigma$ can be performed
explicitly. Again, this results in different expressions for spin
diagonal states ($\sigma_{\nn_1}=\sigma_{\nn_2}$) and spin
off-diagonal states ($\sigma_{\nn_1}\neq\sigma_{\nn_2}$),
\begin{align}
  \tilde{h}_{i,\nn_1\nn_2}^{\tau\sigma\sigma}
  & = \sum_{\nn_3\nn_4} V_{\nn_1\bar{\nn}_2\nn_3\bar{\nn}_4}^{(i)} \big[
    (W_i-H_i)\tilde{\rho}_{\nn_4\nn_3}^{\tau \sigma \sigma}-
    (B_i-M_i)\tilde{\rho}_{\nn_4\nn_3}^{\tau \sigma' \sigma'} \big], \\
  \tilde{h}_{i,\nn_1\nn_2}^{\tau\sigma\sigma'}
  & = \sum_{\nn_3\nn_4} V_{\nn_1\bar{\nn}_2\nn_3\bar{\nn}_4}^{(i)}
  \big(W_i-H_i+B_i-M_i \big) \tilde{\rho}_{\nn_4\nn_3}^{\tau \sigma' \sigma}.
\end{align}

Finally, as before, the summations run over both positive and negative values
of $\Omega$. Therefore, transformation properties of basis states
under time reversal are again used to express the matrix elements of
$\tilde{\rho}$ with negative $\Omega$ in terms of matrix elements with
positive $\Omega$. This results in
\begin{align}
  \label{eq:Delta_sigma}
  \tilde{h}_{i,\nn_1\nn_2}^{\tau\sigma\sigma}
  & = \sum_{\nn_3\nn_4} \Big[
    V_{\nn_1\bar{\nn}_2\nn_3\bar{\nn}_4}^{(i)} (W_i-H_i) -
    V_{\nn_1\bar{\nn}_2\bar{\nn}_3\nn_4}^{(i)} (B_i-M_i) \Big] \tilde{\rho}_{\nn_4\nn_3}^{\tau \sigma \sigma} \nonumber \\
  & \quad \quad - \Big[
    V_{\nn_1\bar{\nn}_2\nn_3\bar{\nn}_4}^{(i)} (B_i-M_i) -
    V_{\nn_1\bar{\nn}_2\bar{\nn}_3\nn_4}^{(i)} (M_i-H_i) \Big] \tilde{\rho}_{\nn_4\nn_3}^{\tau \sigma' \sigma'}, \\
  \label{eq:Delta_sigmap}
  \tilde{h}_{i,\nn_1\nn_2}^{\tau\sigma\sigma'}
  & = \sum_{\nn_3\nn_4} \Big( W_i-H_i+B_i-M_i \Big) \Big(
  V_{\nn_1\bar{\nn}_2\nn_3\bar{\nn}_4}^{(i)} \tilde{\rho}_{\nn_4\nn_3}^{\tau \sigma' \sigma} -
  V_{\nn_1\bar{\nn}_2\bar{\nn}_3\nn_4}^{(i)} \tilde{\rho}_{\nn_4\nn_3}^{\tau \sigma \sigma'} \Big),
\end{align}
where the summations now run only over states with $\Omega>0$. 

{\bf Loop optimization - } The separability of the matrix elements in
axial and polar components is exploited during the calculation of the
HFB fields. The contractions (\ref{eq:GammaDsigma_sigma})-(\ref{eq:GammaEsigma_sigmap}) 
and (\ref{eq:Delta_sigma})-(\ref{eq:Delta_sigmap}) correspond to
contractions of the type
\begin{equation}
  \propto \sum_{\nn_2\nn_4} V_{\nn_1\nn_2\nn_3\nn_4} \rho_{\nn_4\nn_2},
\end{equation}
where the index $\nn_k$ contains only the quantum numbers from the
axially-symmetric harmonic oscillator basis. Naively performing this
summation with $N$ harmonic oscillator shells to calculate all
$\Gamma_{\nn_1\nn_3}$ matrix elements results in an nested loop in
which the number of operations is of the order of $N^{12}$. The number
of operations can be significantly reduced by taking advantage of the
separability of $V_{\nn_1\nn_2\nn_3\nn_4}$ into axial and polar
components. For example, the mean field can be computed as
\begin{align}
  \Gamma_{\nn_1\nn_3} & \propto
  \sum_{\mathclap{n_r^{(2)}\Lambda^{(2)}n_r^{(4)}\Lambda^{(4)}}} 
  V_{n_r^{(1)}\Lambda^{(1)}, n_r^{(2)}\Lambda^{(2)},
    n_r^{(3)}\Lambda^{(3)}, n_r^{(4)}\Lambda^{(4)}}
  \sum_{\mathclap{n_z^{(2)}n_z^{(4)}}} V_{n_z^{(1)}, n_z^{(2)},
    n_z^{(3)}, n_z^{(4)}} \rho_{\nn_4\nn_2},  \\
  & \equiv 
  \sum_{\mathclap{n_r^{(2)}\Lambda^{(2)}n_r^{(4)}\Lambda^{(4)}}} 
  V_{n_r^{(1)}\Lambda^{(1)}, n_r^{(2)}\Lambda^{(2)},
    n_r^{(3)}\Lambda^{(3)}, n_r^{(4)}\Lambda^{(4)}}
  \tensor{Z}_{n_z{(1)}, n_z{(3)}, n_r^{(4)}\Lambda^{(4)},
    n_r^{(2)}\Lambda^{(2)}}.
\end{align}
The calculation of all the $\tensor{Z}$ objects results in a nested
loop where the number of operations increases like $N^8$. The
subsequent contraction of the polar component of the two-body matrix
element with $\tensor{Z}$ requires a loop of order $N^{10}$. This
simple rearrangement in the calculation of $\Gamma$ and $\tilde{h}$ with
$16$ shells reduces the calculation time at each iteration by $97.7\%$.


\subsection{Collective inertia}
\label{subsec:inertia}

The code {\hfbtho} in version \codeversion ~calculates the collective
inertia tensor using either the generator coordinate method (GCM) or the
adiabatic time-dependent Hartree-Fock-Bogolyubov (ATDHFB) approximation.
In the case of the GCM, the calculation is performed using the local
approximation of the Gaussian overlap approximation (GOA), while in the
case of the ATDHFB theory, we use the perturbative cranking approximation;
We give below only a short description of what is implemented in the code;
see \cite{schunck2016} for an recent overview of both methods and
Chapter 10.7 in \cite{ring2000} for a textbook presentation.

{\bf GCM-GOA - } Recall that under the GOA, the many-body Schr\"odinger
equation can be recast into a collective Schr\"odinger-like equation of
the following form,
\be
\hat{\mathcal{H}}_{\text{coll}}(\qvec)=
-\frac{\hbar ^2}{2} \sum_{kl} \frac{\partial}{\partial q_k} B_{kl}(\qvec) \frac{\partial}{\partial q_l}  +  V_{\mathrm{coll}}(\qvec),
\label{eq:GCM-GOA}
\ee
where we note $\qvec = (q_{1}, \dots, q_{N})$ a vector of collective
variables (for instance, each of the $q_i$ would stand for a multipole
moment operator). The collective inertia tensor is $\tensor{B} \equiv B_{kl}(\qvec)$,
and $V_{\mathrm{coll}}(\qvec)$ is the collective potential. It is given by
\be
 V_{\mathrm{coll}}(\qvec)= V(\qvec) - \epsilon_{\mathrm{ZPE}},
\ee
with $V(\qvec)$ the HFB energy for the values $\qvec$ of the collective
variables and $\epsilon_{\mathrm{ZPE}}$ a zero-point energy correction
accounting for the quantum fluctuations of the collective variables.
Neglecting derivative terms, we have
\be
\epsilon_{\mathrm{ZPE}} = \frac{\hbar^{2}}{2} \tensor{\Gamma}\tensor{B},
\ee
where $\tensor{\Gamma}$ is the GCM metric. The term $\tensor{\Gamma}\tensor{B}$ is a
contraction of two rank-2 tensors, that is, $\tensor{\Gamma}\tensor{B} \equiv
\sum_{ab} \Gamma_{ab} B_{ba}$.

In the HFB theory with constraints on the collective variables $\gras{q}$,
one can write the action of the collective momentum $\gras{P} \equiv
-i/\hbar \partial/\partial\qvec$ in terms of quasiparticle creation and
annihilation operators by using the Ring and Schuck theorem \cite{ring1977}.
By invoking linear response theory, the matrix elements of the collective
momentum can also be expressed as a function of the matrix of the collective
variables. This provides a closed-form for the GCM metric, the collective
inertia tensor and the zero-point energy correction, which involve only the
characteristics of the HFB solution at the point $\qvec$ and the QRPA matrix
at the same point. Using the cranking approximation for the QRPA matrix 
provides simple expressions for all these terms. The metric becomes
\be
\tensor{\Gamma} = \frac{1}{2} [\tensor{M}^{(1)}]^{-1} \tensor{M}^{(2)} [\tensor{M}^{(1)}]^{-1},
\ee
with the moments $\tensor{M}^{(K)}$ being
\be
\tensor{M}^{(K)} \equiv M_{\alpha\beta}^{(K)} =
\sum_{\mu<\nu} \frac{\langle\Phi | \hat{Q}^{\dagger}_{\alpha} | \mu\nu\rangle\langle\mu\nu | \hat{Q}_{\beta} | \Phi\rangle}{(E_{\mu}+E_{\nu})^{K}},
\label{eq:moments}
\ee
where $E_{\mu}$ is the quasiparticle energy of quasiparticle $\mu$ and
$|\mu\nu\rangle = \beta^{\dagger}_{\mu}\beta^{\dagger}_{\nu}|\Phi\rangle$ is a 
two-quasiparticle excitation. The term 
$\langle\Phi | \hat{Q}^{\dagger}_{\alpha} | \mu\nu\rangle$ corresponds to 
the block matrix $\tilde{Q}_{\alpha}^{12}$ of the matrix of the operator 
$\hat{Q}_{\alpha}$ in the quasiparticle basis,
\be
\tilde{Q}_{\alpha} = \left(\begin{array}{cc}
\tilde{Q}_{\alpha}^{11} & \tilde{Q}_{\alpha}^{12} \\
\tilde{Q}_{\alpha}^{21} & \tilde{Q}_{\alpha}^{22}
\end{array}\right).
\ee
The collective mass tensor $\tensor{M} = \tensor{B}^{-1}$ reads
\be
\tensor{M}_{\mathrm{GCM}} = 4\tensor{\Gamma} [\tensor{M}^{(1)}] \tensor{\Gamma},
\label{eq:mass_gcm}
\ee
and the zero-point energy can be written
\be
\epsilon_{\mathrm{ZPE}} = \frac{\hbar^{2}}{2} \tensor{\Gamma}\tensor{M}^{-1}.
\label{eq:zpe}
\ee

{\bf ATDHFB - } The ATDHF theory is presented in details in 
\cite{baranger1978}. A quick summary of its extension to include pairing 
correlations is recalled in \cite{schunck2016}. The approximation relies on 
writing the full time-dependent density of the TDHFB theory as the result 
of a unitary transformation $e^{-i\hat{\chi}(t)}$ of a time-dependent, 
time-even generalized density $\mathcal{R}_{0}(t)$, which plays the role of generalized 
coordinate for the collective motion of the nucleus. By subsequently expanding 
the TDHFB energy up to second order in $\hat{\chi}$, one obtains a collective, 
Schr\"odinger-like equation of motion. Further assuming that the time 
dependence of the generalized density $\mathcal{R}_{0}(t)$ is constrained in the collective 
space spanned by the variables $\gras{q} = (q_{1},\dots,q_{N})$, such that
\be
\frac{\partial\mathcal{R}_{0}}{\partial t}
=
\sum_{\alpha} \frac{\partial\mathcal{R}_{0}}{\partial q_{\alpha}}
\frac{\partial q_{\alpha}}{\partial t}
\ee
the collective equation takes nearly the same form as (\ref{eq:GCM-GOA}). The 
main difference is the absence of a zero-point energy correction from the 
collective potential (since the ATDHFB theory is a classical approximation). 
In addition, the expression for the inertia tensor $\tensor{B}$ is slightly 
different from the GCM one. Using both the cranking approximation of the QRPA 
matrix and writing the derivatives 
$\partial\mathcal{R}_{0}/\partial q_{\alpha}$ in terms of the matrix of a 
collective momentum operator, we arrive at
\be
\tensor{M}_{\mathrm{ATDHFB}} 
= \hbar^{2} [\tensor{M}^{(1)}]^{-1} \tensor{M}^{(3)} [\tensor{M}^{(1)}]^{-1},
\label{eq:mass_atdhfb}
\ee
Although conceptually not very well-founded, it is customary to compute a 
zero-point energy correction for the ATDHFB energy by using (\ref{eq:zpe}) 
with the mass tensor
(\ref{eq:mass_atdhfb}).

{\bf Implementation in \textbf{\textsc{hfbtho}} - } The key element of the 
implementation is the calculation of the matrix $\tilde{Q}_{\alpha}^{12}$. Let us 
recall that the matrix of a one-body operator in the doubled single-particle 
basis reads
\be
Q_{\alpha} = \left(\begin{array}{cc}
Q_{\alpha} & 0 \\
0 & -Q_{\alpha}^{*}
\end{array}\right).
\ee
The transformation into the quasiparticle basis to obtain $\tilde{Q}_{\alpha}$ 
is performed with the Bogolyubov transformation
\be
\tilde{Q}_{\alpha} = \mathcal{W}^{\dagger}Q_{\alpha}\mathcal{W},
\ \ \ \ \ 
\mathcal{W} = \left(\begin{array}{cc}
U & V^{*} \\
V & U^{*}
\end{array}\right),
\ee
and yields
\be
\tilde{Q}_{\alpha}^{12} = U^{\dagger}Q_{\alpha}V^{*} - V^{\dagger}Q_{\alpha}^{*}U^{*}.
\ee
At this point, we must consider specifically the structure of the single 
particle basis. The full basis should be complete under time-reversal 
symmetry, that is, for each basis state $|\gras{n}\rangle$, the state 
$|\bar{\gras{n}}\rangle$ is also a basis state. In {\hfbtho}, recall that we 
have $\gras{n} \equiv(n_{r},\Lambda,n_{z},\Sigma)$ and
\be
\phi_{\gras{n}}(\gras{r}\sigma) = \langle\gras{r}\sigma|\gras{n}\rangle 
\equiv \psi_{n_{r}}^{|\Lambda|}(\rho)\psi_{n_{z}}(z)
\frac{e^{i\Lambda\varphi}}{\sqrt{2\pi}}\chi_{\Sigma}(\sigma),
\label{eq:spinor}
\ee
with $\Sigma = \pm 1/2$, $\sigma = 2\Sigma$ and $(\rho,z,\varphi)$ the 
cylindrical coordinates; see \cite{stoitsov2005} for the full definition of 
each wave function. The time-reversed wave function then reads
\be
\phi_{\bar{\gras{n}}}(\gras{r}\sigma) = \langle\gras{r}\sigma|\bar{\gras{n}}\rangle 
= \sigma\psi_{n_{r}}^{|\Lambda|}(\rho)\psi_{n_{z}}(z)
\frac{e^{-i\Lambda\varphi}}{\sqrt{2\pi}}\chi_{-\Sigma}(-\sigma).
\ee
Recall that the HFB spinor for quasiparticle $\mu$ read
\be
\left(\begin{array}{c}
U_{\mu}(\gras{r}\sigma) \\
V_{\mu}(\gras{r}\sigma)
\end{array}\right)
=
\left(\begin{array}{c}
U^{+}_{\mu}(\gras{r}\sigma) \\
V^{+}_{\mu}(\gras{r}\sigma)
\end{array}\right)
e^{i\Lambda^{-}\varphi}\chi_{+1/2}(\sigma)
+
\left(\begin{array}{c}
U^{-}_{\mu}(\gras{r}\sigma) \\
V^{-}_{\mu}(\gras{r}\sigma)
\end{array}\right)
e^{i\Lambda^{+}\varphi}\chi_{-1/2}(\sigma).
\ee
Owing to the fact that time-reversal symmetry is conserved in {\hfbtho}, and 
that the functions $U^{\pm}$ and $V^{\pm}$ are real, the spinor 
\be
\left(\begin{array}{c}
U_{\bar{\mu}}(\gras{r}\sigma) \\
V_{\bar{\mu}}(\gras{r}\sigma)
\end{array}\right)
=
\left(\begin{array}{c}
U^{+}_{\mu}(\gras{r}\sigma) \\
V^{+}_{\mu}(\gras{r}\sigma)
\end{array}\right)
e^{-i\Lambda^{+}\varphi}\chi_{+1/2}(\sigma)
-
\left(\begin{array}{c}
U^{-}_{\mu}(\gras{r}\sigma) \\
V^{-}_{\mu}(\gras{r}\sigma)
\end{array}\right)
e^{-i\Lambda^{-}\varphi}\chi_{-1/2}(\sigma)
\ee
is also a solution of the HFB equation with the same eigenvalue $E_{\mu}$. 
These various relations determine the structure of the matrices of $U$ and 
$V$ in the full s.p. basis. Indeed, we have
\be
U_{\gras{n}\mu} 
= \sum_{\sigma}\int d^{3}\gras{r}\; \phi_{\gras{n}}^{*}(\gras{r}\sigma)U_{\mu}(\gras{r}\sigma),
\ \ \ \ \ 
V_{\gras{n}\mu} 
= \sum_{\sigma}\int d^{3}\gras{r}\; \phi_{\gras{n}}(\gras{r}\sigma)V_{\mu}(\gras{r}\sigma),
\ee
We can then easily show that $U_{\gras{n}\bar{\mu}} = U_{\bar{\gras{n}}\mu} = 0$ and 
$U_{\gras{n}\mu} = U_{\bar{\gras{n}}\bar{\mu}}$, and 
$V_{\gras{n}\mu} = V_{\bar{\gras{n}}\bar{\mu}} = 0$ and 
$V_{\gras{n}\bar{\mu}} = -V_{\bar{\gras{n}}\mu}$. Therefore, the matrices have the 
following generic structure
\be
U \equiv \left(\begin{array}{cc}
U & 0 \\
0 & U
\end{array}\right),
\ \ \ \ 
V \equiv \left(\begin{array}{cc}
 0 & V \\
-V & 0
\end{array}\right).
\ee
Taking into account that the matrices $U$ and $V$ are real and that the 
matrix $Q_{\alpha}$ has the same generic structure as $U$, it then follows 
that 
\be
\tilde{Q}_{\alpha}^{12} = \left(\begin{array}{cc}
 0 & U^{T}Q_{\alpha}V + V^{T}Q_{\alpha}U \\
-U^{T}Q_{\alpha}V - V^{T}Q_{\alpha}U & 0
\end{array}\right).
\ee
In practice, the moments (\ref{eq:moments}) are thus computed through
the formula
\be
M_{\alpha\beta}^{(K)} = 
2\sum_{\mu<\nu} \frac{ (U^{T}Q_{\alpha}V + V^{T}Q_{\alpha}U)_{\mu\nu}(U^{T}Q_{\beta}V + V^{T}Q_{\beta}U)_{\mu\nu}}{(E_{\mu}+E_{\nu})^{K}}
\ee
where the factor 2 comes from the particular structure of the 
$\tilde{Q}_{\alpha}^{12}$ matrix.


\subsection{Fission toolkit}
\label{subsec:fission}

We imported in version {\codeversion} of the code {\hfbtho} a few routines 
implemented in version 2.73 of {\hfodd} to compute the characteristics of 
fission fragments. These include
\begin{itemize}
\item the capability to set the number of particles in the neck by using as a
constraint the expectation value of the Gaussian neck operator
\be
\label{eq:neck}
\hat{Q}_{N} = e^{-(z - z_{N})^{2}/a_{N}^{2}},
\ee
where $z_{N}$ gives the position of the neck (along the $z-$axis of the
intrinsic reference frame) between the two nascent fragments. As in {\hfodd}, it
is defined as the point near the origin of the intrinsic reference frame where
the density is the lowest. The range $a_{N}$ gives the spatial extent of the
neck and is fixed at $a_{N} = 1 $ fm in the code.
\item the charge and mass of each fragment. Expectation values of the multipole
moments are also computed, both with respect to the center of mass of the
fissioning nucleus and that of each of the individual fragments.
\end{itemize}
The various routines are coded in the module \tf{hfbtho\_fission.f90}.
The characteristics of the fission fragments are computed at the convergence
of the HFB iterations by setting the parameter \tv{fission\_fragments = T} in
the namelist \tv{HFBTHO\_FEATURES}. A new namelist for constraints on the
size of the neck called \tv{HFBTHO\_NECK} is available; see section
\ref{subsec:sample} for a description.

We recall that the Gaussian neck operator (\ref{eq:neck}) is purely spatial
and does not depend on spin or isospin degrees of freedom. Its expectation
value is computed in coordinate space on the Gauss-Hermite quadrature grid
used for the $z$-coordinate in \pr{hfbtho}.
\be
\label{eq:neck2}
\langle \hat{Q}_{N} \rangle = \int d^{3}\gras{r}\; \hat{Q}_{N}(\gras{r})\rho(\gras{r}),
\ee
where $\rho(\gras{r})$ is the isoscalar density. The integral is computed by
Gauss quadrature. Since \pr{hfbtho} works in cylindrical coordinates
$(\rho,z,\varphi)$ and uses the stretched coordinates $\xi_{z} = b_z z$ and
$\eta = b_{\perp}^{2}\rho^2$. The code tabulates the value of the Gaussian
neck operator on the Gauss-Laguerre and Gauss-Hermite meshes,
$Q_{N}(\xi_{h},\eta_{\ell})$. The expectation value is then given by
\be
\langle \hat{Q}_{N} \rangle =
\sum_{\ell=1}^{N_{\text{Lag}}} \sum_{h=1}^{N_{\text{Her}}}
\omega_{\ell}\omega_{h} Q_{N}(\xi_{h},\eta_{\ell}) \rho(\xi_{h},\eta_{\ell})
\ee
with $\omega_{h}$, $\omega_{\ell}$ and $\xi_{h}, \eta_{\ell}$ the weights and
nodes of the Gauss-Hermite and Gauss-Laguerre quadrature, respectively.

The expectation value of the neck operator can be set by imposing a linear
constraint in the HFB equation. The Lagrange parameter controlling this
constraint is readjusted based on the cranking approximation of the QRPA
matrix; see section 2.4 in \cite{stoitsov2013} and section 2.3 in
\cite{schunck2017} for details about the method. This technique requires
the matrix elements of the operator in the HO basis. In \pr{hfbtho}, the
matrix elements are computed directly by first defining the value of the
Gaussian neck operator on the Gauss-Laguerre and Gauss-Hermite meshes and
then computing the matrix elements in each $\Omega$ block as
\begin{eqnarray}
\langle n_{r}\Lambda n_{z}\sigma|\hat{Q}_{N}| n'_{r}\Lambda' n'_{z}\sigma'\rangle
& = & \int d^{3}\gras{r}\; \phi_{\nn}^{*}(\gras{r},\sigma)
\hat{Q}_{N}(\gras{r}) \phi_{\nn'}(\gras{r},\sigma') \\
& = & \delta_{\sigma\sigma'}\delta_{n_{r}n'_{r}}\delta_{\Lambda\Lambda'}
\sum_{h=1}^{N_{\text{Her}}}
w_{h} \psi_{n_{z}}(z_{h}) Q_{N}(z_h) \psi_{n'_{z}}(z_{h})
\end{eqnarray}
where $\phi_{\nn}^{*}(\gras{r},\sigma)$ are the basis spinor (\ref{eq:spinor}); 
see \cite{stoitsov2005} for additional definitions.

Given the position of the neck, we calculate the mass of the two fission
fragment classically, that is, by integrating the density up to the
neck position,
\be
A_{1} = \int d\varphi \int \rho^2 d\rho \int_{-\infty}^{z_{N}} dz \rho(\gras{r}),
\ \ \ \ 
A_{2} = A - A_{1}.
\ee
A similar formula applies to the charge of the fragments. Since the
integral over $z$ does not extend to $+\infty$, Gauss-Hermite quadrature
cannot be used. In our implementation, we interpolate the value of the
density along the $z$-axis using spline functions, and compute the
integral over $z$ using the Simpsons's rule on a mesh of 1000 points.
Although more accurate integration schemes could be implemented, our
choice is numerically accurate enough to within the 4th digit of the
value of $A_1$.

To compute the deformations of the fission fragments in their intrinsic frame, we
first compute the positions of the centers of mass of each fragment,
\be
z_{1} = \int d\varphi \int \rho^2d\rho \int_{-\infty}^{z_{N}} zdz,
\ \ \ \ \ 
z_{2} = \int d\varphi \int \rho^2d\rho \int_{z_{N}}^{+\infty} zdz.
\ee
Recall that the multipole moments operators are defined in spherical 
coordinates by
\be
\hat{Q}_{\lambda\mu}(r,\theta,\varphi) = r^{\lambda}Y_{\lambda\mu}(\theta,\varphi).
\ee
where $Y_{\lambda\mu}(\theta,\varphi)$ are the spherical harmonics.
Since \pr{hfbtho} assumes axial symmetry, only $\mu=0$ multipole moments
are non-zero. It is easy to show that they can be expressed in terms of 
$\rho$ and $z$ only. The expectation value of the multipole moment of 
order $\lambda$ in the intrinsic frame of the  fission fragment $1$ is 
thus given by
\be
\langle Q_{\lambda 0} \rangle_{1} =
\int d\varphi \int \rho^{2}d\rho \int_{-\infty}^{z_{N}} dz\; 
\hat{Q}_{\lambda 0}(\rho,z-z_{1}) \rho(\gras{r}).
\ee
The expression for the second fragment is obtained by integrating over $z$
from $z_{N}$ to $+\infty$. As with the charge and mass numbers of the
fission fragments, we choose to interpolate the values of the density along
the $z$-axis with Spline functions and carry out all integrals over $z$
with the Simpson's 3/8 rule.


\subsection{Regularization of zero-range pairing force}
\label{subsec:regu}

The code {\hfbtho} now implements the regularization of zero-range pairing
forces according to the original idea introduced by Bulgac in \cite{bulgac2002}.
Our implementation follows most closely \cite{bulgac2002,bennaceur2005}. It is
limited to the case of functionals of the local pairing density
$\tilde{\rho}(\gras{r})$ only. Moreover, following arguments given in
\cite{bennaceur2005}, and explicit benchmarks of \cite{borycki2006}, the 
Hartree-Fock field resulting from the variations of the Fermi momentum $k_{F}$ 
and cut-off momentum $k_{c}$ with respect to the density are not implemented in 
version {\codeversion}.

Under these conditions, the pairing energy is written as an integral of a
functional of the local pairing density,
\be
E_{\mathrm{pair}} =
\sum_{q=n,p} \int d^{3}\gras{r}\; g_{q}(\gras{r})\tilde{\rho}_{q}^{2}(\gras{r}),
\ee
and the local pairing field is simply
\be
\tilde{U}_{q}(\gras{r}) = 2g_{q}(\gras{r})\tilde{\rho}_{q}(\gras{r}).
\ee
The regularization procedure consists in computing an effective pairing
strength function $g_{\text{eff}}(\gras{r})$ such that these expressions simply
become
\be
E_{\mathrm{pair}} =
\sum_{q=n,p} \int d^{3}\gras{r}\; g_{\text{eff},q}(\gras{r})\tilde{\rho}_{q}^{2}(\gras{r}),
\ \ \ 
\tilde{U}_{q}(\gras{r}) = 2g_{\text{eff},q}(\gras{r})\tilde{\rho}_{q}(\gras{r}).
\ee

For each type of particles $q$ (protons or neutrons), the position-dependent 
effective pairing strength $g_{\text{eff}}(\gras{r})$ is given by the table 
below, based on the real or imaginary nature of the two cut-off momenta 
$\kc$ and $\lc$ and the Fermi momentum $\kF$:
\begin{equation*}
\begin{array}{c|c}
\displaystyle \Gape[0pt][9pt]{1/g_{\text{eff}}(\gras{r})} & \text{condition} \\
\hline\\
\displaystyle\Gape[0pt][9pt]{
\begin{array}{r}
\displaystyle
\frac{1}{g(\gras{r})} - \frac{\Mstar \kc}{2\pi^2\hbar^{2}}
\left( 1 - \frac{\kF}{2\kc} \log\frac{\kc + \kF}{\kc - \kF} \quad\right. 
\smallskip\\
\displaystyle
\left. - \frac{\kF}{2\lc} \log\frac{\kF + \lc}{\kF - \lc}\right)
\end{array}
}
& \ell^{2}_{\text{cut}}\geq 0, k^{2}_{\text{cut}}\geq 0, k_{F}^{2}\geq 0 \\ 
\displaystyle\Gape[0pt][9pt]{
\frac{1}{g(\gras{r})} - \frac{\Mstar \kc}{2\pi^2\hbar^{2}}
\left( 1 - \frac{\kF}{2\kc} \log\frac{\kc + \kF}{\kc - \kF} \right) 
}
& \ell^{2}_{\text{cut}}< 0, k^{2}_{\text{cut}}\geq 0, k_{F}^{2}\geq 0 \\
\displaystyle
\Gape[0pt][9pt]{
\frac{1}{g(\gras{r})} - \frac{\Mstar \kc}{2\pi^2\hbar^{2}}
\left( 1 + \frac{|\kF|}{\kc} \arctan\frac{|\kc|}{\kc} \right)
}
& \ell^{2}_{\text{cut}}< 0, k^{2}_{\text{cut}}\geq 0, k_{F}^{2}< 0 \\
\displaystyle
\frac{1}{g(\gras{r})}
& \ell^{2}_{\text{cut}}< 0, k^{2}_{\text{cut}}<0, k_{F}^{2}< 0 \\
\end{array}
\end{equation*}
In these expressions, the scalar effective mass for particle $q$ is defined as
\be
\Mstar =
\frac{\hbar^{2}}{2m}
+ (C_{0}^{\rho\tau} - C_{1}^{\rho\tau})\rho_{0}(\gras{r})
+ 2C_{1}^{\rho\tau}\rho_{q}(\gras{r}),
\ee
where $C_{t}^{\rho\tau}$, $t=0,1$ are the isoscalar and isovector coupling
constants of the kinetic density part of the Skyrme functional, $\rho_{0}$ is
the isoscalar density and $\rho_{q}$ the density for particle $q$. The Fermi
momentum $\kF$ is given by
\be
\kF = \frac{\sqrt{2\Mstar}}{\hbar}\sqrt{\mu - U(\gras{r})},
\ee
where $\mu$ is the Fermi energy and $U(\gras{r})$ is the self-consistent mean
field potential; cf. Eq.(18) in \cite{stoitsov2005}. The cutoff momenta read
\be
\begin{array}{l}
\kc = \displaystyle\frac{\sqrt{2\Mstar}}{\hbar}\sqrt{\mu + \epsilon_{\text{cut}} - U(\gras{r})},
\medskip\\
\lc = \displaystyle\frac{\sqrt{2\Mstar}}{\hbar}\sqrt{\mu - \epsilon_{\text{cut}} - U(\gras{r})}
\end{array}
\ee
with $\epsilon_{\text{cut}}$ the cut-off in quasiparticle energies (relative
to the Fermi energy), $\epsilon_{\text{cut}} = E_{\text{max}} - \mu$. Recall
that $E_{\text{max}}$ is an input parameter of \pr{hfbtho} defined in the
namelist under the keyword \tf{pairing\_cutoff}.


\subsection{Localization functions}
\label{subsec:loc}

The code \pr{hfbtho} in version {\codeversion} calculates the nucleon spatial localization with fixed spin and isospin in subroutine \tf{hfbtho\_localization.f90}. The spatial localization was originally introduced in atomic and molecular physics to characterize chemical bond structures in electron systems \cite{becke1990,savin1997,scemama2004,kohout2004,burnus2005,poater2005}, and then was applied to identify cluster structures in light nuclei \cite{reinhard2011} and fragment formation in fissioning heavy nuclei \cite{zhang2016}.

The definition of spatial localization is based on the probability of finding a nucleon within a distance $\delta$ from a given nucleon at position $\gras{r}$ with the same spin $\sigma$ ($=\uparrow$ or $\downarrow$) and isospin $q$ ($=n$ or $p$). By applying a Taylor expansion with respect to $\delta$, this probability can be written as
\begin{equation}\label{eqn:probability}
R_{q\sigma}(\gras{r},\delta)\approx\frac{1}{3}\left(\tau_{q\sigma}-\frac{1}{4}\frac{|\gras{\nabla}\rho_{q\sigma}|^2}{\rho_{q\sigma}}-\frac{\gras{j}_{q\sigma}^2}{\rho_{q\sigma}}\right)\delta^2+\mathcal{O}(\delta^3),
\end{equation}
where $\rho_{q\sigma}$, $\tau_{q\sigma}$, $\gras{j}_{q\sigma}$, and $\gras{\nabla}\rho_{q\sigma}$ are the particle density, kinetic energy density, current density, and density gradient, respectively. Through the canonical HFB orbits $\psi_{\alpha}$, they can be expressed as
\begin{subequations}
\begin{eqnarray}
\rho_{q\sigma}(\gras{r})&=&\sum_{\alpha\in q}v^2_{\alpha}|\psi_\alpha(\gras{r}\sigma)|^2,\\
\tau_{q\sigma}(\gras{r})&=&\sum_{\alpha\in q}v^2_{\alpha}|\gras{\nabla}\psi_\alpha(\gras{r}\sigma)|^2,\\
\gras{j}_{q\sigma}(\gras{r})&=&\sum_{\alpha\in q}v^2_{\alpha}\mathrm{Im}[\psi^*_\alpha(\gras{r}\sigma)\gras{\nabla}\psi_\alpha(\gras{r}\sigma)],
\\
\gras{\nabla}\rho_{q\sigma}(\gras{r})&=&2\sum_{\alpha\in q}v^2_{\alpha}\mathrm{Re}[\psi^*_\alpha(\gras{r}\sigma)\gras{\nabla}\psi_\alpha(\gras{r}\sigma)],
\end{eqnarray}
\end{subequations}
with $v^2$ being the canonical occupation probability. Thus, the expression in the parentheses of (\ref{eqn:probability}) can serve as a localization measure. Since the spatial localization and $R_{q\sigma}$ are in a reverse relationship, we can introduce a dimensionless expression for the spatial localization by normalizing with the Thomas-Fermi kinetic energy density $\tau_{q\sigma}=\frac{3}{5}(6\pi^2)^{2/3}\rho_{q\sigma}^{5/3}$,
\begin{equation} \label{eqn:localization}
\mathcal{C}_{q\sigma}(\gras{r})=\left[1+\left(\frac{\tau_{q\sigma}\rho_{q\sigma}-{1\over 4}|\gras{\nabla}\rho_{q\sigma}|^2-\gras{j}^2_{q\sigma}}{\rho_{q\sigma}\tau^\mathrm{TF}_{q\sigma}}\right)^2\right]^{-1}.
\end{equation}
In \pr{hfbtho}, time reversal symmetry is always conserved, therefore, $\gras{j}_{q\sigma}$ vanishes.


\subsection{MPI capabilities}
\label{subsec:mpi}

Two new large scale production modes have been implemented in version 
\codeversion, {\tt DO\_MASSTABLE} and {\tt DRIP\_LINES}. To activate any of 
these two new modes, the corresponding pre-processor directive must be set 
to {\tt 1} in the {\tt makefile} before compilation. To activate MPI for 
any of these two modes the pre-processor directive {\tt USE\_MPI} must be 
set to {\tt 1}.

The {\tt DO\_MASSTABLE} mode allows performing, serially or in parallel, a 
list of {\hfbtho} calculations with different number of protons, number of 
neutrons, expectation value for $\hat{Q}_2$ and basis deformation. The list 
of calculations is read from the {\tt hfbtho\_MASSTABLE.dat} file. The 
second line of the file must indicate the number of calculations to be made 
and the list must start at the fourth line. An example file is provided with 
the current version. Once all the calculations have been finished a summary 
of the obtained binding energies is sent to the standard output. An output 
file with a unique label is produced for every calculation in the mass table.

\begin{figure}[!ht]
\center
\includegraphics[width=0.8\linewidth]{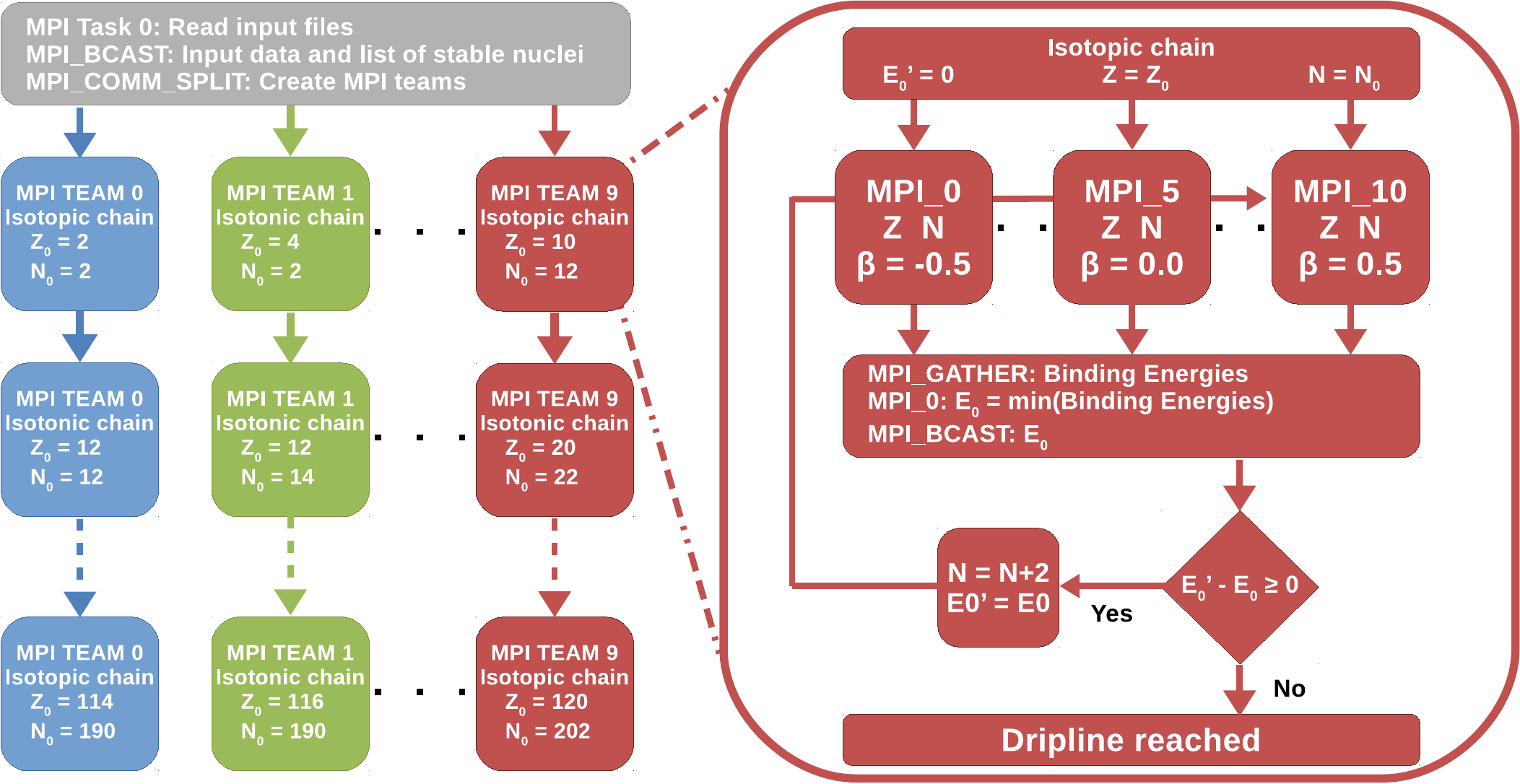}
\caption{(color online) Flow chart outlining the distribution of HFB
  calculations during execution on {\tt DRIP\_LINES} mode. The left
  side of the chart shows the distribution of all the isotopic and
  isotonic chains between the different MPI teams. The right side
  shows the calculation of each chain which includes, the distribution
  of HFB calculations with different deformations, the process to
  calculate the ground state binding energy of each nucleus and the
  criterion to determine the position of the corresponding dripline.}
\label{fig:driplines_flow_chart}
\end{figure}

The {\tt DRIP\_LINES} mode performs a full scan of all even-even nuclei in 
the nuclear chart by looking for the two-neutron and two-proton drip-lines. 
This mode requires {\tt USE\_MPI = 1} and the number of MPI 
tasks must be a multiple of 11. In future versions the second requirement 
could be removed although significant load unbalances might occur as a result. 
In this mode, MPI teams are formed to run over isotopic or isotonic chains 
calculating the binding energy of every nucleus with 11 different deformations. 
The list of isotopic and isotonic chains to be explored is read from the 
{\tt hfbtho\_STABLELINE.dat}. For every nucleus the deformation with the 
lowest binding energy is considered the ground-state. After the ground-state 
is found the MPI team will move into the next even-even nuclei on the isotopic 
(or isotonic) chain. The drip-line is considered to be reached when the 
two-neutron (-proton) separation energy becomes negative. The two-neutron
separation energy is defined as
\begin{equation}
  S_{\rm 2n}(Z,N) = B(Z,N-2)-B(Z,N),
\end{equation}
and an analogous expression is used for the two-proton separation
energy.  This process is presented schematically in figure
\ref{fig:driplines_flow_chart}.

For bookkeeping purposes and to facilitate any subsequent review of
every HFB calculation, an output file with a unique label is created
for each calculation. Furthermore, each MPI team creates a log file
named {\tt TeamTableXXX.dat} that indicates the nuclei and deformation
corresponding to each label for the output files.


\subsection{Other changes}
\label{subsec:other}

{\bf Input/output - } We have changed the way the code records output on disk.
In the past two releases of {\hfbtho}, the name of the file where the program 
was recording data, in binary format, contained the number of protons and 
neutrons explicitly. In addition, the program was recording the matrix elements 
of the HFB matrix, which is dependent on the definition of the HO basis used 
in the run. This format was not especially adapted to large-scale calculations 
of, e.g., full mass tables or potential energy surfaces (PES), where it is often 
advantageous to use a solution at a given point to initialize the calculation 
of another point -- the two points having possibly different basis 
characteristics. Therefore, we have instead adopted the same convention as in 
{\hfodd}, where the program records the values of the mean field and pairing
field in the quadrature mesh. In practice, the same quadrature mesh can be used
in a full PES calculation without significant loss of precision. As a result
of this format change, we also modified the name of the binary file to give it
the generic name {\tt hfbtho\_output.hel}. The new format is much more flexible
and enables to initialize a new calculation even when the HO basis, the nucleus,
the interaction, etc., are different.\\

\noindent{\bf Charge radius - } We have adopted the following definition of the
charge radius
\be
r_{\text{ch.}} = \sqrt{\langle r_{\text{p}}\rangle^2 + \langle R^2_{p} \rangle
+ \frac{N}{Z}\langle R^{2}_{n}\rangle + \frac{3}{4M_{p}^{2}}}
\ee
where $\langle r_{\text{p}}\rangle^2$ is the expectation value on the HFB
vacuum of the proton radius; $\langle R^2_{p}\rangle  = 0.769$ fm$^{2}$ is the
proton charge radius; $\langle R^{2}_{n}\rangle = -0.1161$ fm$^{2}$ is the neutron
charge radius taken from \cite{yao2006}; $3/4M_{p}^{2} = 0.033$ fm$^{2}$ is the
Darwin-Foldy term from \cite{friar1997}.\\

\noindent{\bf Kick-off mode - } We have improved the handling of what was called
kick-off constraints in the previous version of {\hfbtho}. These constraints
are imposed only for the first 10 iterations and are subsequently released.
This was introduced originally to push the HFB solution towards the correct
minimum. In version {\codeversion}, we allow the use of
regular and kick-off constraints side-by-side. For example, the user can set a constraint
on the expectation value of $\hat{Q}_{10}$ and $\hat{Q}_{30}$ while using
the kick-off mode for $\hat{Q}_{20}$.


\section{Benchmarks and Accuracy}
\label{sec:benchmarks}

We give in this section several numerical benchmarks for the new features 
added in the current version of {\hfbtho}. Input data files corresponding to 
the benchmarks for the Gogny force, the collective inertia and the fission 
toolkit are included in the code release. For the regularization method, 
we provide the input data file corresponding to the value of the cutoff of 
$E_{\text{pair}} = 80$ MeV.

\setcounter{mysubsubsection}{0}


\subsection{Gogny force}
\label{subsec:bench_gogny}

The implementation of the Gogny interaction has been benchmarked against the
symmetry-unrestricted DFT solver {\hfodd}. We provide two main benchmarks:
\begin{itemize}
\item An unconstrained HFB calculation in $^{120}$Sn using a deformed
basis characterized by a spherical-equivalent oscillator length of
$b_0 = 2.1339661000$ fm, an axial deformation $\beta = 0.4$, a maximum
number of shells $N_{\rm max}=20$ and a maximum number of states of
$N_{\rm states}=1771$. Such a large basis allowed probing the numerical
accuracy of the two-body potential matrix elements with a large number of
shells.
\item A constrained HFB calculation in $^{152}$Dy using a deformed
basis characterized by a spherical-equivalent oscillator length of
$b_0 = 2.12100948454$ fm, an axial deformation $\beta = 0.3$, a maximum
number of shells $N_{\rm max}=14$ and a maximum number of states of
$N_{\rm states}=500$. We imposed only a constraint of
$\langle \hat{Q}_{20} \rangle = 20$ b on the expectation value of the
axial quadrupole moment.
\end{itemize}
Both benchmarks were made without including the Coulomb interaction in
order to test only the implementation of the D1S Gogny functional: the
treatment of the  Coulomb potential is slightly different in the two codes,
which can  lead to small differences of the order of the keV; see
\cite{schunck2012,stoitsov2013}. The benchmarks are shown in
Tables~\ref{tab:Gogny120Sn}-\ref{tab:Gogny152Dy}.

\begin{table}[!ht]
\center
 \caption{\label{tab:Gogny120Sn} Benchmark of the DFT solvers {\hfbtho}
   and {\hfodd} for a Hartree-Fock-Bogolyubov calculation in $^{120}$Sn
   with the D1S Gogny functional. We choose on purpose a deformed basis
   with $N_{\rm max}=20$ and $\beta = 0.4$. The spherical-equivalent of
   the oscillator length is $b_0=2.1339661000$ fm ($b_{\perp} = 2.0035166$
   fm and $b_{z} = 2.4208988$ fm). The deformation of the initial solution
   is set to the same value as the basis deformation, that is $\beta_0 = 0.4$.}
 \begin{tabular}{ l l l  }
   & {\hfbtho} & {\hfodd} \\
 \hline \noalign{\smallskip}
 $E_{\rm tot}$ (MeV)                &           -1385.1669{\htho 90} &           -1385.1669{\hodd 81}\\
 $E_{\rm kin}^{\rm (n)}$ (MeV)      & \phantom{-}1416.3633{\htho 21} & \phantom{-}1416.3633{\hodd 01}\\
 $E_{\rm kin}^{\rm (p)}$ (MeV)      & \phantom{-1}888.5274{\htho 64} & \phantom{-1}888.5274{\hodd 52}\\
 $E_{\rm Vol}$ (MeV)                & \phantom{-}4004.862{\htho 420} & \phantom{-}4004.862{\hodd 313}  \\
 $E_{\rm SO}$ (MeV)                 & \phantom{11}-72.659{\htho 299} & \phantom{11}-72.659{\hodd 315}\\
 $E_{\rm Gogny}^{\rm (dir)}$ (MeV)  &           -7274.909{\htho 836} &           -7274.909{\hodd 691}\\
 $E_{\rm Gogny}^{\rm (exc)}$ (MeV)  & \phantom{1}-326.6363{\htho 58} & \phantom{1}-326.6363{\hodd 90}\\
 $E_{\rm Gogny}^{\rm (pair)}$ (MeV) & \phantom{11}-20.714{\htho 701} & \phantom{11}-20.714{\hodd 651}\\
 $r_{\rm rms}^{\rm (n)}$ (fm)       & \phantom{-111}4.626410         & \phantom{-111}4.626410\\
 $r_{\rm rms}^{\rm (n)}$ (fm)       & \phantom{-111}4.438686         & \phantom{-111}4.438686
 \end{tabular}
\end{table}

\begin{table}[!ht]
\center
 \caption{\label{tab:Gogny152Dy} Benchmark of the DFT solvers {\hfbtho}
   and {\hfodd} for a constrained Hartree-Fock-Bogolyubov calculation in
   $^{152}$Dy with the D1S Gogny functional. We choose a deformed basis
   with $N_{\rm max}=14$, $\beta = 0.3$. The spherical-equivalent of
   the oscillator length is $b_0=2.12100948454$ fm ($b_{\perp} = 2.0230038$
   fm and $b_{z} = 2.3314948$ fm). The deformation of the initial solution
   is set to the same value as the basis deformation, that is $\beta_0 = 0.4$.
   We impose a constraint of $\langle \hat{Q}_{20} \rangle = 20$ b.}
 \begin{tabular}{ l l l }
   & {\hfbtho} & {\hfodd} \\
 \hline \noalign{\smallskip}
 $E_{\rm tot}$ (MeV)                &           -1816.2879{\htho 80} &           -1816.2879{\hodd 78}\\
 $E_{\rm kin}^{\rm (n)}$ (MeV)      & \phantom{-}1742.87{\htho 2954} & \phantom{-}1742.87{\hodd 3034}\\
 $E_{\rm kin}^{\rm (p)}$ (MeV)      & \phantom{-}1199.2861{\htho 39} & \phantom{-}1199.2861{\hodd 69}\\
 $E_{\rm Vol}$ (MeV)                & \phantom{-}5224.217{\htho 198} & \phantom{-}5224.217{\hodd 576} \\
 $E_{\rm SO}$ (MeV)                 & \phantom{11}-90.2434{\htho 11} & \phantom{11}-90.2434{\hodd 58}\\
 $E_{\rm Gogny}^{\rm (dir)}$ (MeV)  &           -9463.410{\htho 240} &           -9463.410{\hodd 712}\\
 $E_{\rm Gogny}^{\rm (exc)}$ (MeV)  & \phantom{1}-411.011{\htho 819} & \phantom{1}-411.011{\hodd 793}\\
 $E_{\rm Gogny}^{\rm (pair)}$ (MeV) & \phantom{11}-17.998{\htho 801} & \phantom{11}-17.998{\hodd 793}\\
 $r_{\rm rms}^{\rm (n)}$ (fm)       & \phantom{-111}5.103611         & \phantom{-111}5.103611\\
 $r_{\rm rms}^{\rm (n)}$ (fm)       & \phantom{-111}4.959125         & \phantom{-111}4.959125
 \end{tabular}
\end{table}

We see that enforcing strictly identical HO basis in both codes lead to
agreement between kinetic energies of less than 100 eV, and of less than
500 eV on the direct part of the Gogny interaction energy. Note that we
use the recently published version 2.73y of {\hfodd}, where the
implementation of the Gogny force was also benchmarked against a spherical
code.


\subsection{Large-amplitude collective motion}
\label{subsec:bench_lacm}

We show in table \ref{tab:inertia} the results for the components of the 
collective mass tensor, both in the GCM approximation of (\ref{eq:mass_gcm}) 
and in the ATDHFB approximation of (\ref{eq:mass_atdhfb}), as well as the 
zero-point energy correction. Calculations were performed in $^{240}$Pu for 
the SkM* parametrization of the Skyrme potential and a surface-volume pairing 
force characterized by $V_0^{(n)} = -265.2500$ MeV, $V_0^{(p)} = -340.0625$ 
MeV and a cut-off on quasiparticle energies of $E_{\mathrm{max}} = 60$ MeV; 
see \cite{schunck2014} for additional details. The basis was characterized 
by $b_{0} = 2.5$ fm, $\beta = 0.9$, and $N_{\text{max}} = 20$. An additional 
cutoff on the total number of states was imposed with $N_{\text{states}} = 800$. 
We imposed the constraints, $\langle \hat{Q}_{20}\rangle = 320$ b and 
$\langle \hat{Q}_{30}\rangle = 25$ b$^{3/2}$, which correspond to a region not 
far from the scission point. Calculations were verified against an unpublished 
branch of version 2.73 of {\hfodd}. The collective inertia module was also
benchmarked separately against the triaxial Gogny code of L. Robledo and 
results on each component agree up to the third digit.

\begin{table}[!ht]
\center
 \caption{\label{tab:inertia} Benchmark of the collective inertia mass tensor 
 and zero-point energy corrections at the perturbative cranking approximation; 
 see text for details.}
 \begin{tabular}{ l l l l }
   & & {\hfbtho} & {\hfodd} \\
 \hline \noalign{\smallskip}
 \multirow{4}{4em}{GCM}    & M$_{22}$ (MeV$^{-1}$b$^{-2}$)   & \phantom{-}0.2016.10$^{-2}$         &  \phantom{-}0.2016.10$^{-2}$ \\
                           & M$_{32}$ (MeV$^{-1}$b$^{-5/2}$) & -0.5787.10$^{-3}$                   &            -0.5787.10$^{-3}$ \\
                           & M$_{33}$ (MeV$^{-1}$b$^{-3}$)   & \phantom{-}0.3878.10$^{-2}$         &  \phantom{-}0.3878.10$^{-2}$ \\
                           & $\epsilon_{\text{ZPE}}$ (MeV)   & \phantom{-}2.1785                   &  \phantom{-}2.1784   \medskip\\
 \multirow{4}{4em}{ATDHFB} & M$_{22}$ (MeV$^{-1}$b$^{-2}$)   & \phantom{-}0.282{\htho 2}.10$^{-2}$ &  \phantom{-}0.282{\hodd 6}.10$^{-2}$ \\
                           & M$_{32}$ (MeV$^{-1}$b$^{-5/2}$) & -0.8667.10$^{-3}$                   &            -0.8667.10$^{-3}$ \\
                           & M$_{33}$ (MeV$^{-1}$b$^{-3}$)   & \phantom{-}0.484{\htho 8}.10$^{-2}$ &  \phantom{-}0.484{\hodd 9}.10$^{-2}$ \\
                           & $\epsilon_{\text{ZPE}}$ (MeV)   & \phantom{-}1.6819                   &  \phantom{-}1.6819
 \end{tabular}
\end{table}

Next, we illustrate the calculation of fission fragments properties. The 
characteristics of the run were identical as for the collective inertia. 
Although the discrepancy between the two codes is a little larger than 
for previous benchmarks, especially for multipole moments, note that this 
is simply a consequence of performing integration over $]-\infty,z_{N}]$ 
or $[z_{N}, +\infty[$, for which there is no exact quadrature rule. 
Numerical differences between the codes are also much smaller than, 
e.g., basis truncation errors.

\begin{table}[!ht]
\center
 \caption{\label{tab:fission} Benchmark of the characteristics of the fission 
 fragments at the point $\langle q_{20}\rangle = 320$ b and 
 $\langle q_{30}\rangle = 25 $ b$^{3/2}$. We list the dimensionless coordinate 
 of the neck $\xi_{N} = z_{N}/b_{z}$. $z_{\mathrm{CM}1}$ (resp. $z_{\mathrm{CM}2}$) 
 give the position of the center of mass of fragment 1 (resp., fragment 2) and is 
 given in fermis; see text for additional details.}
 \begin{tabular}{ l l l l }
   & {\hfbtho} & {\hfodd} \\
 \hline \noalign{\smallskip}
 $\xi_{N}$               & \phantom{-11}0.6418         & \phantom{-11}0.6418  \\
 q$_{N}$                 & \phantom{-11}8.8442         & \phantom{-11}8.8442  \\
 z$_{\mathrm{CM}1}$ (fm) & \phantom{11}-6.301{\htho 4} & \phantom{11}-6.301{\hodd 3} \\
 z$_{\mathrm{CM}2}$ (fm) & \phantom{-11}9.1059         & \phantom{-11}9.1059  \\
 Z$_{1}$                 & \phantom{-1}56.005{\htho 5} & \phantom{-1}56.005{\hodd 3} \\
 Z$_{2}$                 & \phantom{-1}37.994{\htho 5} & \phantom{-1}37.994{\hodd 7} \\
 A$_{1}$                 & \phantom{-}141.8{\htho 508} & \phantom{-}141.8{\hodd 480} \\
 A$_{2}$                 & \phantom{-1}98.1{\htho 492} & \phantom{-1}98.1{\hodd 397} \\
 q$_{20,1}$ (b)          & \phantom{-1}29.2{\htho 889} & \phantom{-1}29.2{\hodd 793} \\
 q$_{20,2}$ (b)          & \phantom{-1}15.2{\htho 483} & \phantom{-1}15.2{\hodd 234} \\
 q$_{30,1}$ (b$^{3/2})$  & \phantom{-11}1.6{\htho 485} & \phantom{-11}1.6{\hodd 521} \\
 q$_{30,2}$ (b$^{3/2})$  & \phantom{11}-0.3{\htho 751} & \phantom{11}-0.3{\hodd 832} \\
 \end{tabular}
\end{table}


\subsection{Regularization of zero-range pairing force}
\label{subsec:bench_regu}

We have tested our implementation of the pairing regularization by focusing 
on the nucleus $^{120}$Sn. We use a full spherical basis of $N=20$ shells 
(hence $N_{\text{states}} = 1771$) and oscillator length 
$b_{0} = 2.03901407814296$ fm. We solved the HFB equation for the SLy4 
parametrization of the Skyrme functional. In the pairing channel, we use 
a surface-volume pairing force characterized by the same pairing strength 
for protons and neutrons, $V_{0} = -370.2$ MeV. This should give us conditions 
very similar to those reported in \cite{bennaceur2005}; see column 4 of Table 
1 and line 3 of Table 2. 

\begin{figure}[!ht]
\center
\includegraphics[width=0.7\linewidth]{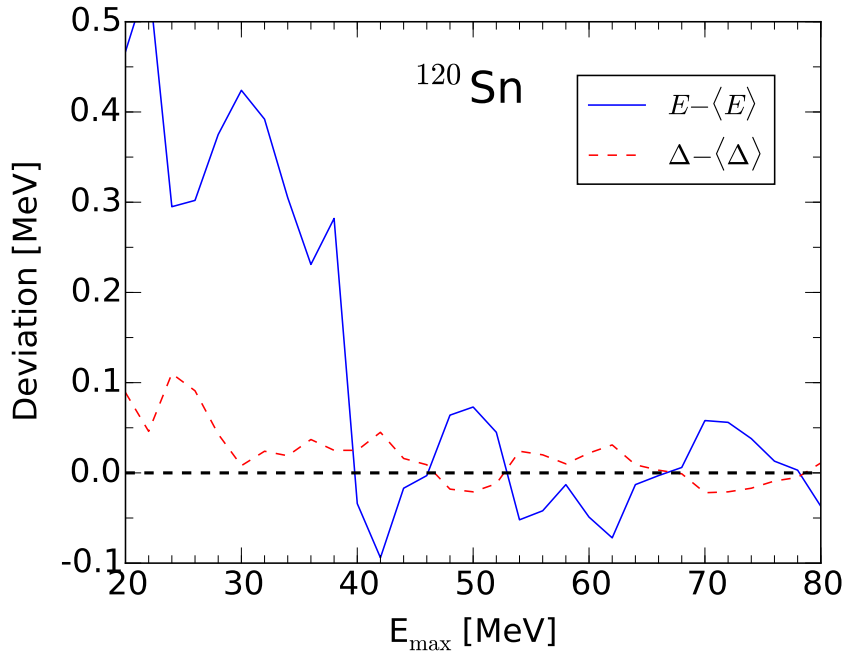}
\caption{(color online) Difference between the total energy $E$ and its mean 
value $\langle E \rangle$ (solid line) and between the average neutron gap 
and its mean value (dashed line) as a function of the maximum energy 
$E_{\text{max}}$ for $^{120}$Sn. The mean values are calculated over the 
interval $60 \leq E_{\text{max}} \leq 80 $ MeV.
}
\label{fig:regularization}
\end{figure}

We performed the calculation for several values of the cut-off energy, here 
denoted $E_{\text{max}}$. As in \cite{bennaceur2005}, we show in figure 
\ref{fig:regularization} the deviation between the total energy and neutron 
pairing gaps from their respective mean values. The mean values were computed 
in the range $60 \leq E_{\text{max}} \leq 80 $ MeV of cut-off energy, and are 
equal to $\langle E \rangle = -1018.396$ MeV and 
$\langle \Delta \rangle = 1.251$ MeV. As expected, these values are a little 
different from calculations performed in coordinate space. In particular, the 
136 keV difference in energy is compatible with systematics shown, e.g., in 
\cite{schunck2015-c}.

Overall, the variations of the energy and of the pairing gaps are about 3-4 
times larger than for the {\hfbrad} solver. This result was expected and is 
discussed in more details in \cite{borycki2006}. Note that we could not 
reproduce the {\hfbtho} results of \cite{borycki2006} for three reasons: the 
version of the code used back then did not properly take into account the 
pairing rearrangement energy, resulting in small errors of typically 10-20 keV 
on the total energy; the treatment of the direct Coulomb energy was numerically 
inaccurate, as discussed extensively in \cite{stoitsov2013}; finally, the value 
of the pairing strength used in the studies of \cite{borycki2006} is not listed.


\subsection{Multi-threading}
\label{subsec:openmp}

\begin{figure}[!ht]
\center
\includegraphics[width=0.57\linewidth]{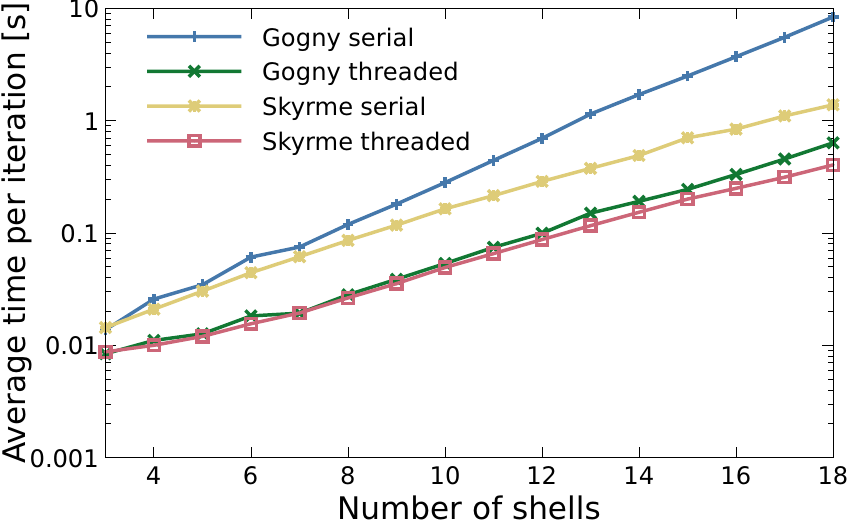}
\ 
\includegraphics[width=0.37\linewidth]{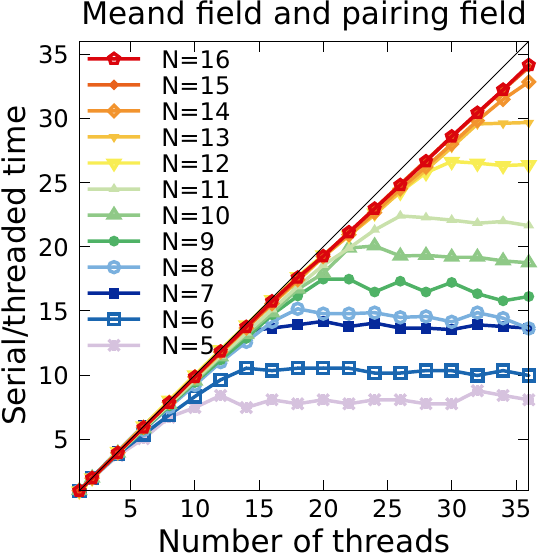}
\caption{(color online) Left: average time per HFB iteration (in log scale) 
as a function of the number of shells of the HO basis for the serial and 
multi-threaded version of the code, for both Skyrme and Gogny functionals. 
36 OpenMP threads are used. Right: ratio between serial and threaded 
execution time for the calculation of the mean field and pairing field as 
a function of the number of threads. The diagonal black line represents 
a perfect parallelization.
}
\label{fig:openmp}
\end{figure}

Finally, we illustrate the advantage of multi-threading for HFB calculations. 
The left panel of figure \ref{fig:openmp} shows the average time per iteration 
of a spherical HFB calculation as a function of the size of the basis. Here, 
calculations are done in a full spherical basis, hence the number of states 
is proportional to the number of shells $N$ according to $N_{\mathrm{states}} = 
(N+1)(N+2)(N+3)/6$. The figure shows results obtained in $^{50}$Cr with either 
the SLy6 parametrization of the Skyrme potential or the D1S parametrization 
of the Gogny force. In the case of the Skyrme force, a surface-volume 
interaction was used in the pairing channel. Threaded calculations were done 
with 36 threads and run on the Quartz supercomputer at LLNL. As expected, 
OpenMP multi-threading can drastically accelerate code execution, by about an 
order of magnitude for the Gogny force and a factor 5 for Skyrme forces. 
Interestingly, multi-threaded HFB calculations with the Gogny force are only 
about twice slower than multi-threaded HFB calculations with the Skyrme force. 

To better understand the impact of shared memory parallelism, the right panel 
of figure \ref{fig:openmp} shows the ratio between serial and threaded 
execution time for the calculation of the finite-range contribution to the 
mean field and pairing field only. 
For instance, a ratio of 10 for 10 threads indicates that the threaded code 
is 10 times faster, which is perfect strong scaling. For the largest basis 
considered here ($N_{\mathrm{max}}=16$), the code achieves an impressive 
95\% of maximum strong scaling. Conversely, performance saturates for small 
basis sizes whenever the number of threads is (approximately) twice the 
number of shells in the basis.


\section{Input data file}
\label{sec:input}

The input data file format remains similar to version 2.00d and only contains 
a few additional namelists.


\subsection{Sample input file}
\label{subsec:sample}

\begin{verbatim}
&HFBTHO_GENERAL
 number_of_shells = 10, oscillator_length = -1.0, basis_deformation = 0.0,
 proton_number = 24, neutron_number = 26, type_of_calculation = 1 /
&HFBTHO_ITERATIONS
 number_iterations = 100, accuracy = 1.E-5, restart_file = -1 /
&HFBTHO_INITIAL
 beta2_deformation = 0.0, beta4_deformation = 0.0 /
&HFBTHO_FUNCTIONAL
 functional = 'SLY4', add_initial_pairing = F, type_of_coulomb = 2 /
&HFBTHO_PAIRING
 user_pairing = F, vpair_n = -300.0, vpair_p = -300.0,
 pairing_cutoff = 60.0, pairing_feature = 0.5 /
&HFBTHO_CONSTRAINTS
 lambda_values = 1, 2, 3, 4, 5, 6, 7, 8,
 lambda_active = 0, 0, 0, 0, 0, 0, 0, 0,
 expectation_values = 0.0, 0.0, 0.0, 0.0, 0.0, 0.0, 0.0, 0.0 /
&HFBTHO_BLOCKING
 proton_blocking = 0, 0, 0, 0, 0, neutron_blocking = 0, 0, 0, 0, 0 /
&HFBTHO_PROJECTION
 switch_to_THO = 0, projection_is_on = 0,
 gauge_points = 1, delta_Z = 0, delta_N = 0 /
&HFBTHO_TEMPERATURE
 set_temperature = F, temperature = 0.0 /
&HFBTHO_FEATURES
 collective_inertia = F, fission_fragments = F, pairing_regularization = F,
 localization_functions = F /
&HFBTHO_NECK
 use_constrain = F, neck_value = 0.5 /
&HFBTHO_DEBUG
 number_Gauss = 40, number_Laguerre = 40, number_Legendre = 80,
 compatibility_HFODD = F, number_states = 500, force_parity = T,
 print_time = 0 /
\end{verbatim}


\subsection{Description of input data}
\label{subsec:description}

We now define the new classes of input used in version {\codeversion}.

\key{HFBTHO\_INITIAL}

\noindent$\bullet$ {\tt beta2\_deformation = 0.0}: The axial quadrupole
deformation $\beta_{2}$ of the Woods-Saxon potential used to initialize
the calculation. Default: 0.0\\

\noindent$\bullet$ {\tt beta4\_deformation = 0.0}: The axial hexadecapole
deformation $\beta_{4}$ of the Woods-Saxon potential used to initialize
the calculation. Default: 0.0\\

\key{HFBTHO\_FEATURES}

\noindent$\bullet$ {\tt collective\_inertia = F}: Logical switch that
activates the calculation of the collective inertia tensor and zero-point
energy corrections in the ATDHFB and GCM approach (both within the
cranking approximation. The calculation of these quantities is only
carried out at convergence. Default: F\\

\noindent$\bullet$ {\tt fission\_fragments = F}: Logical switch that
activates the calculation of the characteristics of the fission fragments.
Note that in order to perform such a calculation, the code must not
enforce the conservation of parity (through the input {\tt force\_parity}).
If the code detects that {\tt force\_parity = T}, {\em it will override
its value to False} to guarantee that parity can be broken during the
self-consistent loop. As a result of this condition, iterations may be
noticeably slower, even though fission fragment properties are only
computed at convergence. Default: F\\

\noindent$\bullet$ {\tt pairing\_regularization = F}: Logical switch that
activates the regularization of the pairing strength for zero-range,
density-dependent pairing forces. The value of this keyword has no
effect on calculations with the finite-range Gogny force. Default: F\\

\noindent$\bullet$ {\tt localization\_functions = F}: Logical switch that
activates the calculation of the localization functions at convergence.
Default: F\\

\key{HFBTHO\_NECK}

\noindent$\bullet$ {\tt use\_constrain = F}: Logical switch that activates
the constraint on the size of the neck. Default: F\\

\noindent$\bullet$ {\tt neck\_value = 0.5}: Value of the constraint on
the expectation value of the Gaussian neck operator. Default: 0.5\\


\section{Program \textsc{hfbtho}}
\label{sec:program}


\subsection{Structure of the code}
\label{subsec:code}

Compared with version 2.00d, the program \pr{hfbtho} has been broken into
several different files implementing specific Fortran modules:
\begin{itemize}
\item {\tt hfbtho\_bessel.f90}: modified Bessel functions of order 0 and 1;
\item {\tt hfbtho\_collective.f90}: collective inertia tensor and zero-point
energy correction at the perturbative cranking approximation;
\item {\tt hfbtho\_elliptic\_integrals.f90}: complete elliptic integral of
the second kind;
\item {\tt hfbtho\_fission.f90}: charge, mass and axial multipole moments of
fission fragments and Gaussian neck operator;
\item {\tt hfbtho\_gauss.f90}: Gauss-Hermite, -Laguerre, and -Legendre
quadrature meshes;
\item {\tt hfbtho\_gogny.f90}: matrix elements of the Gogny force;
\item {\tt hfbtho\_large\_scale.f90}: MPI parallel interface for mass table
calculations;
\item {\tt hfbtho\_linear\_algebra.f90}: various mathematical routines;
\item {\tt hfbtho\_localization.f90}: spatial localization functions;
\item {\tt hfbtho\_main.f90}: main calling program;
\item {\tt hfbtho\_multipole\_moments.f90}: expectation value and matrix
elements of axial multipole moments;
\item {\tt hfbtho\_read\_functional.f90}: user-defined parameters of a
generalized Skyrme-like energy functional;
\item {\tt hfbtho\_solver.f90}: HFB self-consistent loop, expectation value,
HF and pairing field and input/output routines;
\item {\tt hfbtho\_tho.f90}: transformed harmonic oscillator basis;
\item {\tt hfbtho\_unedf.f90}: parameterizations of the Skyrme and Gogny 
functionals and density-dependent coupling constants and
fields of generalized Skyrme energy functionals;
\item {\tt hfbtho\_utilities.f90}: definition of integer and real types and
constants;
\item {\tt hfbtho\_variables.f90}: list of global variables used throughout
the code;
\item {\tt hfbtho\_version.f90}: version number and old history;
\end{itemize}
The programming language of most of the code is Fortran 95, while legacy code
is still written, in part or totally, in Fortran 90 and Fortran 77. The code
\pr{{\hfbtho}} requires an implementation of the BLAS and LAPACK libraries to
function correctly. Shared memory parallelism is available via OpenMP pragmas. 

The new version also comes with a built-in Doxygen documentation. To benefit 
from this feature, the user should install the doxygen software available at 
\url{www.doxygen.org}. The documentation is built by typing
\begin{center}
{\tt make doc }
\end{center}
By default, Doxygen generates only an on-line HTML documentation. The main 
page is located at {\tt ./doc/html/index.html}. A PDF documentation can also 
be generated by going into {\tt ./doc/latex} and typing
\begin{center}
{\tt make}
\end{center}
The PDF file is named {\tt refman.pdf}.


\subsection{Running the code}
\label{subsec:run}

The program ships with a Makefile that is preset for a number of
Fortran compilers. The user should choose the compiler and set the path
for the BLAS and LAPACK libraries. Assuming an executable named
{\tt hfbtho\_main} and a Linux system, execution is started by typing
\begin{center}
{\tt ./hfbtho\_main < /dev/null >\& hfbtho\_main.out }
\end{center}
The program will attempt to read the file named {\tt hfbtho\_NAMELIST.dat}
in the current directory. The user is in charge of ensuring this file is
present, readable, and has the proper format. The code will automatically
generate a binary file of the form named {\tt hfbtho\_output.hel}.

As shown in section \ref{sec:benchmarks}, HFB calculations are greatly 
accelerated when OpenMP multi-threading is activated. However, the user 
should keep in mind that this requires setting additional environment 
variables. In Linux/Unix machines, the default stack size is not large 
enough to run the code and must be increased. This can be achieved by 
instructions such as
\begin{center}
\parbox{0.7\textwidth}{
\noindent {\tt ulimit -s unlimited}\\
\noindent {\tt export OMP\_STACKSIZE=32M }
}
\end{center}

The value of {\tt ulimit}  defines the amount of stack size for the main 
OpenMP thread. OpenMP supports control over the stack size limit of all 
additional threads via the environment variable {\tt OMP\_STACKSIZE}. The 
value given above should be sufficient for all applications. Note that this 
value does not affect the stack size of the main thread. For completeness, 
note that the GNU OpenMP run-time (libgomp) recognizes the non-standard 
environment variable {\tt GOMP\_STACKSIZE}. If set it overrides the value of 
{\tt OMP\_STACKSIZE}. Finally, the Intel OpenMP run-time also recognizes the 
non-standard environment variable {\tt KMP\_STACKSIZE}. If set it overrides 
the value of both {\tt OMP\_STACKSIZE} and {\tt GOMP\_STACKSIZE}.


\section{Acknowledgments}
\label{sec:acknowledgments}

\bigskip
We are very grateful to Luis Robledo for having helped us with the benchmark 
of the collective inertia tensor. We also thank Karim Bennaceur for providing 
us with the latest version of {\hfbrad} to test the regularization of the 
pairing channel, and Tomas Rodriguez for providing access to his axial code 
to test the Gogny force.
Support for this work was partly provided through Scientific Discovery
through Advanced Computing (SciDAC) program funded by U.S. Department of
Energy, Office of Science, Advanced Scientific Computing Research and
Nuclear Physics. It was partly performed under the auspices of the US
Department of Energy by the Lawrence Livermore National Laboratory under
Contract DE-AC52-07NA27344 (code release number: LLNL-CODE-573953, document
release number: LLNL-JRNL-587360). An award of computer time was
provided by the Innovative and Novel Computational Impact on Theory and
Experiment (INCITE) program. This research used resources of the Oak Ridge
Leadership Computing Facility located in the Oak Ridge National Laboratory,
which is supported by the Office of Science of the Department of Energy
under Contract DE-AC05-00OR22725. It also used resources of the National
Energy Research Scientific Computing Center, which is supported by the
Office of Science of the U.S. Department of Energy under Contract
No. DE-AC02-05CH11231. We also acknowledge ``Fusion,''a 320-node cluster
operated by the Laboratory Computing Resource Center at Argonne National
Laboratory.


\bibliography{zotero_output,biblio,books}
\bibliographystyle{cpc}

\end{document}